%% file: paper.tex
\documentclass[12pt]{article}
\usepackage{jheppub}

\pdfoutput=1

\usepackage[table]{xcolor}
\usepackage{amsmath,array,amsfonts,graphicx,wrapfig,lscape,float,mathtools,multirow,longtable,setspace}
\usepackage{physics}
\usepackage{stackrel}
\usepackage[all]{xy}
\usepackage{subcaption}
\usepackage{url}
\usepackage{cancel}
\usepackage{array}

\input{pref}

\definecolor{darkspringgreen}{rgb}{0.09, 0.45, 0.27}
\definecolor{forestgreen}{rgb}{0.13, 0.55, 0.13}
\definecolor{blue2}{RGB}{65,128,255}
\definecolor{yellow2}{rgb}{0.98, 0.80, 0.20}

\usepackage{array}

\usepackage{array}
\usepackage{tikz}
\usepackage{tikz-3dplot}
\usetikzlibrary{decorations.pathreplacing,decorations.markings,arrows.meta,shapes.misc}
\tikzset{cross/.style={cross out, draw=black, minimum size=2*(#1-\pgflinewidth), inner sep=0pt, outer sep=0pt},
cross/.default={1pt}}

\newcolumntype{C}[1]{>{\centering\let\newline\\\arraybackslash\hspace{0pt}}m{#1}}

\newcommand{\floor}[1]{ \left\lfloor{ #1} \right\rfloor}

\renewcommand{\d}{\partial }

\newcommand{\beas}{\begin{equation} \begin{aligned}} \newcommand{\eeas}{\end{aligned} \end{equation}}

\newcommand{\arrowHeadPosition}{0.75}

\title{Calabi-Yau Products: Graded Quivers for General Toric Calabi-Yaus} 

\author[a,b,c]{Sebasti\'an Franco,} 
\author[d]{Azeem Hasan}

\affiliation[a]{
Physics Department, The City College of the CUNY \\
160 Convent Avenue, New York, NY 10031, USA}

\affiliation[b]{Physics Program and $^c$Initiative for the Theoretical Sciences \\
The Graduate School and University Center, The City University of New York  \\
365 Fifth Avenue, New York NY 10016, USA}

\affiliation[d]{Dipartimento di Matematica e Fisica Ennio De Giorgi \\
Universit\`{a} del Salento \& INFN, Via Arnesano, 73100 Lecce, Italy}

\emailAdd{sfranco@ccny.cuny.edu}
\emailAdd{azeem.hasan@le.infn.it}

\abstract{The open string sector of the topological B-model on CY $(m+2)$-folds is described by $m$-graded quivers with superpotentials.  This correspondence generalizes the connection between CY $(m+2)$-folds and gauge theories on the worldvolume of D$(5-2m)$-branes for $m=0,\ldots,3$ to arbitrary $m$. In this paper we introduce the Calabi-Yau product, a new algorithm that starting from the known quiver theories for a pair of toric CY$_{m+2}$ and CY$_{n+2}$ produces the quiver theory for a related CY$_{m+n+3}$. This method significantly supersedes existing ones, enabling the simple determination of quiver theories for geometries that were previously out of practical reach.
}

\begin{document}

\maketitle

\section{Introduction}

The engineering of gauge theories in different dimensions by means of branes probing Calabi-Yau (CY) singularities in string and M-theory has received considerable attention. Among its multiple applications, this approach: provides a way to construct interesting gauge theories and study their dynamics and dualities, is a framework for local model building \cite{Aldazabal:2000sa,Berenstein:2001nk,Verlinde:2005jr,Buican:2006sn} and it is at the heart of the gauge/gravity correspondence \cite{Maldacena:1997re,Gubser:1998bc,Witten:1998qj}.

The well-known connection between CY $(m+2)$-folds and gauge theories on the worldvolume of D$(5-2m)$-branes for $m=0,\ldots,3$ (see e.g. \cite{Morrison:1998cs,Beasley:1999uz,Feng:2000mi,Beasley:2001zp,Feng:2001xr,Feng:2001bn,Feng:2002zw,Wijnholt:2002qz,Benvenuti:2004dy,Franco:2005rj,Benvenuti:2005ja,Franco:2005sm,Butti:2005sw} for the widely studied case of D3-branes on CY 3-folds) can be extended to arbitrary $m$ in terms of the topological B-model. In this context, the open string sector of the B-model on CY $(m+2)$-folds is described by $m$-graded quivers with superpotentials (see \cite{Aspinwall:2008jk,lam2014calabi,Franco:2017lpa,Closset:2018axq} and references therein). 

This correspondence is particularly well understood in the case of toric CYs. For $m=1$, {\it brane tilings} (a.k.a. dimer models), significantly simplify the map between CY 3-folds and $4d$ $\mathcal{N}=1$ gauge theories \cite{Hanany:2005ve,Franco:2005rj,Franco:2005sm}. Progress in this area has considerably accelerated in recent years, initially fueled by a desire to develop brane constructions for lower dimensional gauge theories \cite{Franco:2015tna,Franco:2015tya,Franco:2016nwv,Franco:2016qxh,Franco:2016tcm,Franco:2017cjj}. Lately, the scope of these investigations expanded to developing tools for toric CYs of arbitrary dimension. These efforts culminated in \cite{Franco:2019bmx} with the introduction of {\it $m$-dimers}, which fully encode the $m$-graded quivers with superpotentials associated to toric CY $(m+2)$-folds and streamline the connection between quivers and geometry.

The $m$-dimers associated to specific geometries can be determined via a variety of traditional approaches, such as partial resolution and mirror symmetry, which have been extended to general $m$ \cite{Franco:2015tna}. Despite the considerable simplifications brought by $m$-dimers, their determination can sometimes become practically challenging and additional tools are desirable. Examples of such methods include {\it orbifold reduction} \cite{Franco:2016fxm} and {\it $3d$ printing} \cite{Franco:2018qsc} which were originally developed in the context of CY 4-folds but can be applied more broadly \cite{Closset:2018axq}.

In this paper we introduce a substantially more powerful approach, which we denote {\it Calabi-Yau product}. This algorithm starts from the known quiver theories\footnote{Throughout this paper, we will use the term quiver theory to indicate the combination of a quiver and its superpotential.} for a pair of toric CY$_{m+2}$ and CY$_{n+2}$ and produces the quiver theory for a related CY$_{m+n+3}$. In doing so, it enables the computation of quiver theories that were previously out of practical reach.

This paper is organized as follows. \sref{section_graded_quivers} presents a review of $m$-graded quivers. \sref{section_product_geometry} introduces the basics of the CY product, in particular the input data for the construction and how the parent geometries give rise to the product geometry. \sref{section_periodic_quiver} explains how to construct the periodic quiver for the product theory. \sref{section_superpotential} discusses the superpotential. The construction is illustrated in \sref{section_examples} with explicit examples. \sref{section_relation_to_other_constructions} considers the relation between the CY product and other constructions. We conclude and present ideas for future work in \sref{section_conclusions}. Additional details are provided in two appendices.

\section{A Brief Review of $m$-Graded Quiver Theories}

\label{section_graded_quivers}

In order to make our presentation self-contained, in this section we present a brief review $m$-graded quivers and their dualities. We refer the interested reader to \cite{Franco:2017lpa,Closset:2018axq,Franco:2019bmx} for further details.

Given an integer $m \geq 0$, an $m$-graded quiver is a quiver with a grading for every arrow $\Phi_{ij}$ by a {\it quiver degree}:
\beq
|\Phi_{ij}| \in \{ 0, 1, \cdots, m\}~.
\eeq
Every node $i$ corresponds to a unitary ``gauge group" $U(N_i)$. Arrows connecting nodes correspond to bifundamental or adjoint ``fields".

The conjugate of every arrow $\Phi_{ij}$ has the opposite orientation and degree $m-|\Phi_{ij}|$:
\beq\label{Phi opp intro}
\overline{\Phi}_{ji}^{(m-c)}\equiv \overline{(\Phi_{ij}^{(c)})}~,
\eeq
where we use a superindex in parenthesis to explicitly indicate the degree of the corresponding arrow, i.e. $|\Phi_{ij}^{(c)}|=c$. 

The integer $m$ determines the possible degrees, i.e. the different types of fields, which can be restricted to the range:
\beq\label{arrows fields}
\Phi_{ij}^{(c)} \; : i \longrightarrow j~, \qquad c=0, 1, \cdots, n_c-1~, \qquad n_c \equiv    \floor{m+2\over 2}~,
\eeq
since other degrees can be obtained by conjugation. We refer to degree 0 fields as {\it chiral fields}.

Graded quivers for $m=0,1,2,3$ describe $d=6,4,2,0$ supersymmetric gauge theories with $2^{3-m}$ supercharges, respectively. Different degrees correspond to different types of superfields. These theories can be engineered in terms of Type IIB D$(5-2m)$-branes probing CY $(m+2)$-folds.

\paragraph{Superpotential.}

Graded quivers admit {\it superpotentials}, which are linear combinations of gauge invariant terms of degree $m-1$:
\be
W= W(\Phi)~,\qquad\qquad  |W|= m-1~.
\label{superpotential_degree}
\ee
Gauge invariant terms correspond to closed oriented cycles in the quiver, which may require conjugation of some of the fields.

\paragraph{Kontsevich bracket condition.}
The superpotential must also satisfy
\beq
\{W,W \}=0 ~.
\label{superpotential_Kontsevich}
\eeq 
Here $\{ f, g \}$ denotes the Kontsevich bracket, which is defined as follows   
\be
\{ f, g \}= \sum_\Phi \left( {\d f\over \d \Phi}{\d g\over \d\overline{\Phi}} +(-1)^{(|f|+1)|\overline{\Phi}|+(|g|+1)|\Phi|+ |\Phi||\overline{\Phi}|+1} {\d f\over \d \overline{\Phi}}{\d g\over \d \Phi} \right)~.
\ee

\subsection{The Toric Case}
 
  \label{section_toric_quivers}     
 
The CY$_{m+2}$ associated to an $m$-graded quiver arises as its {\it classical moduli space} which, generalizing the standard notion for $m\leq 3$, is defined as the center of the Jacobian algebra with respect to fields of degree $m-1$ \cite{Franco:2017lpa}. Namely, it is obtained by imposing the relations:
\beq
{\partial W \over \partial \Phi^{(m-1)}}=0~, \ \ \ \ \ \forall \, \Phi^{(m-1)}
\label{relations}
\eeq
plus gauge invariance. Since the superpotential has degree $m-1$, the terms that contribute to the relations in \eref{relations} are of the general form $\Phi^{(m-1)} J(\Phi^{(0)})$, with $J(\Phi^{(0)})$ a holomorphic function of chiral fields. We will refer to such terms as $J$-terms. The relations \eref{relations} therefore comprise only chiral fields.

\paragraph{Toric superpotential.}

Every toric CY$_{m+2}$ has at least one {\it toric phase}, which is a quiver theory satisfying the following properties. First, the ranks for all nodes can be equal. In addition, the superpotential of a toric phase has a special structure, which is referred to as the {\it toric condition} \cite{Franco:2019bmx}. The toric condition implies that every field of degree $m-1$ appears in exactly two superpotential terms, with opposite signs. Namely,
\beq
W= \Phi^{(m-1)}_a J_a^+(\Phi^{(0)})- \Phi_a^{(m-1)} J_a^-(\Phi^{(0)}) + \ldots~,
\label{general_toric_W}
\eeq
where dots stand for terms that do not contain $\Phi_a^{(m-1)}$. The relations \eref{relations} then take the form:
\beq
J_a^+(\Phi^{(0)}) = J_a^-(\Phi^{(0)}) ~.
\label{general_toric_J}
\eeq

Due to this special structure, toric phases can be encoded in $m$-dimers or, equivalently, by periodic quivers on $\mathbb{T}^{m+1}$ \cite{Franco:2019bmx}.

\paragraph{Generalized perfect matchings.}
    
We define a {\it generalized perfect matching}, or perfect matching for short, $p$ as a collection of fields satisfying:
    \begin{itemize}
        \item[{\bf 1)}] $p$ contains precisely one field from each term in $W$.
        \item[{\bf 2)}] For every field $\Phi$ in the quiver, either $\Phi$ or $\bar{\Phi}$ is in $p$. 
    \end{itemize}

Perfect matchings provide variables that automatically satisfy the relations \eref{general_toric_J}. Therefore, there is a one-to-one correspondence between them and GLSM fields in the toric description of the CY$_{m+2}$. Perfect matchings indeed substantially simplify the determination of the toric diagram (see \cite{Franco:2019bmx} for details).
 
Since for every field a perfect matching contains either the field or its conjugate, a perfect matching determines a {\it polarization} of the quiver.  We define polarization as a choice of orientation for every field in the quiver, i.e. a choice of what we regard as the original field and its conjugate. In what follows, we will adopt a convention for defining the polarization such that, given a perfect matching, we orient the fields in the quiver such that the fields in the perfect matching are the only ones that appear conjugated in the superpotential.\footnote{Notice that while every perfect matching defines a polarization, not every polarization corresponds to a perfect matching. For a quiver with $N_f$ fields, there are $2^{N_f}$ possible polarizations, arising from the two choices of orientation for every field.} This choice of polarization implies that the corresponding perfect matching consists of the conjugates of all the fields in the quiver.

\subsection{Dualities}
 \label{subsec_dualities}

$m$-graded quivers admit order $(m+1)$ mutations. For $m\leq 3$, they correspond to the dualities of the corresponding gauge theories: no duality for $6d$ $\mathcal{N}=(0,1)$, Seiberg duality for $4d$ $\mathcal{N}=1$ \cite{Seiberg:1994pq}, triality for $2d$ $\mathcal{N}=(0,2)$ \cite{Gadde:2013lxa} and quadrality for $0d$ $\mathcal{N}=1$ \cite{Franco:2016tcm}. Interestingly, these mutations generalize these dualities to $m>3$. We refer the reader to \cite{Franco:2017lpa,Closset:2018axq} for detailed discussions on the transformation of quiver theories under mutations.

\subsection{Generalized anomaly cancellation}

Under a mutation at a node $\star$, its rank transform as:
\beq
N'_\star = N_0 - N_\star ~,
\eeq
where $N_0$ is the total number of incoming chiral fields. Invariance of the ranks under $m+1$ consecutive mutations of the same node leads to the {\it generalized anomaly cancellation} conditions. 
For odd $m$, these conditions are given by:
\beq
\sum_j N_j \sum_{c=0}^{n_c-1} (-1)^c \left(\CN(\Phi_{ji}^{(c)})-\CN(\Phi_{ij}^{(c)})\right)=0~, \qquad \forall i~, \qquad {\rm if}\;\; m \in 2\mathbb{Z}+1~,
\label{anomaly_odd}
\ee
with $\CN(\Phi_{ij}^{(c)})$ denotes the number of arrows from $i$ to $j$ of degree $c$. For every $i$, the sum over $j$ runs over all nodes in the quiver (including $i$), and $n_c$ is given by \eref{arrows fields}.

For even $m$, the conditions become
\be
\sum_j N_j \sum_{c=0}^{n_c-1}(-1)^c \left(\CN(\Phi_{ji}^{(c)})+\CN(\Phi_{ij}^{(c)})\right)=2N_i~, \qquad \forall i~, \qquad {\rm if}\;\; m \in 2\mathbb{Z}~.
\label{anomaly_even}
\ee
For $m=0,1,2,3$, these conditions reproduce the cancellation of non-abelian anomalies in the corresponding $d=6,4,2,0$ gauge theories.

\section{Product of Toric Calabi-Yaus: the Geometry}

\label{section_product_geometry}

In this paper we will introduce the CY product. Before explaining the details of this novel algorithm, let us discuss its main ingredients and basics of the resulting geometry.

\paragraph{Initial data.}

The input for this procedure is given by:
\begin{itemize}
\item An $m$-graded quiver theory $P$ for a toric phase associated with a toric Calabi-Yau $(m+2)$-fold $\mathrm{CY}_{m+2}$. The toric diagram $T_{\mathrm{CY}_{m+2}}$ is an $(m+1)$-dimensional convex polytope consisting of points $u_{i}$. We also pick a perfect matching $p$ of $P$, which corresponds to the point $u_{0}$ of $T_{\mathrm{CY}_{m+2}}$.

\item An $n$-graded quiver theory $Q$ for a toric phase associated with a toric Calabi-Yau $(n+2)$-fold $\mathrm{CY}_{n+2}$. The toric diagram $T_{\mathrm{CY}_{n+2}}$ is an $(n+1)$-dimensional convex polytope consisting of points $v_{i}$ in it. We also pick a perfect matching $q$ of $Q$, which corresponds to the point $v_{0}$ of $T_{\mathrm{CY}_{n+2}}$.
\end{itemize}

\paragraph{The product geometry.}
  
\begin{figure}[ht]
	\centering
	\includegraphics[width=13.5cm]{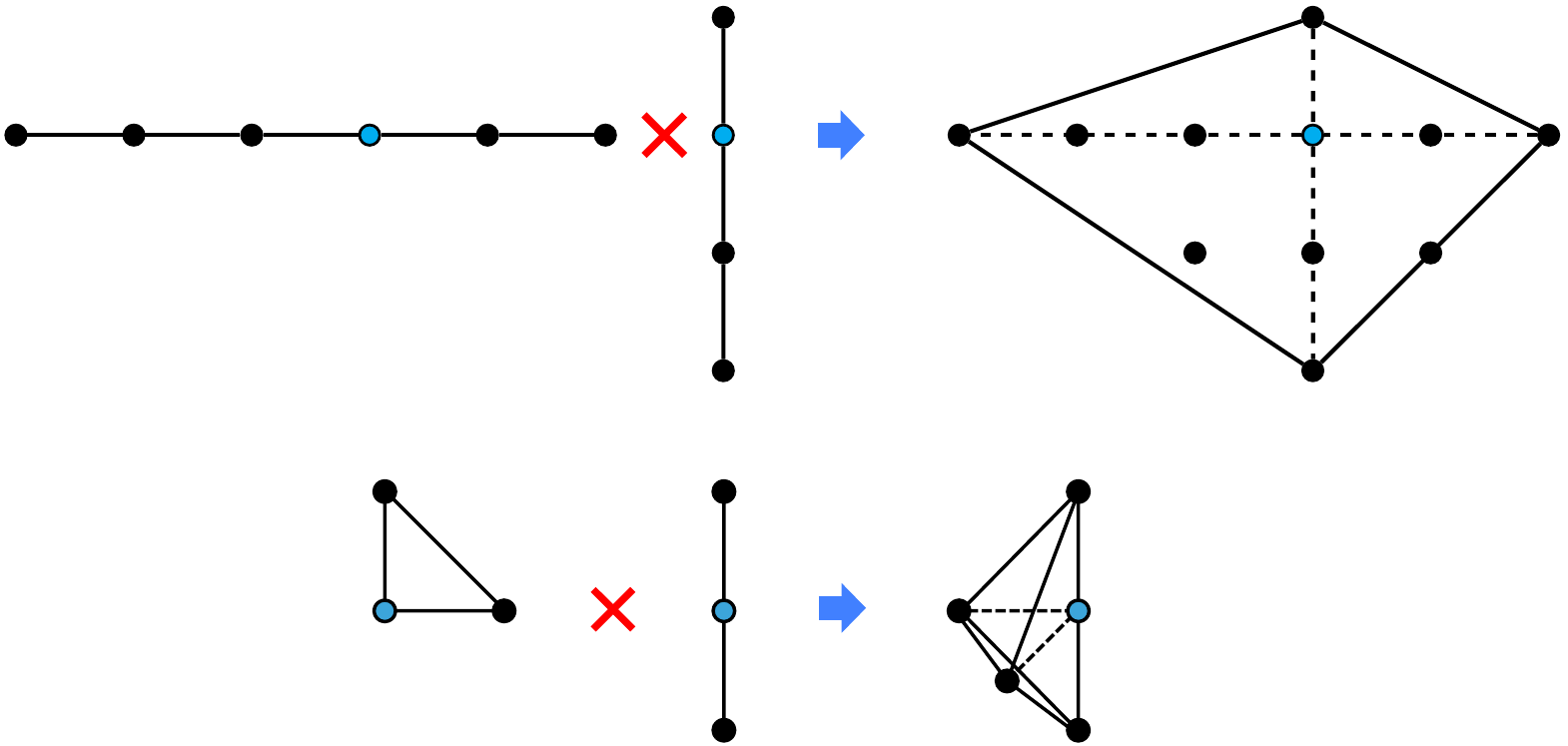}
\caption{Two examples of the action of the Calabi-Yau product on toric diagrams. The first line is an example of CY$_2\times$CY$_2=$CY$_3$. The second line is CY$_3\times$CY$_2=$CY$_4$.}
	\label{examples_product_toric_diagrams}
\end{figure}

The output of this algorithm is an $(m+n+1)$-graded quiver theory that we will call $P_{p}\times Q_{q}$. This theory is a toric phase for the $(m+n+3)$-dimensional toric Calabi-Yau $\mathrm{CY}_{m+n+3}$ whose toric diagram $T_{\mathrm{CY}_{m+n+3}}$ is the convex hull of points
\beq
        \{(u_{i},v_{0})|u_{i} \in T_{\mathrm{CY}_{m+2}}\} \cup \{(u_{0},v_{i})|v_{i} \in T_{\mathrm{CY}_{n+2}}\}  
        \label{product_toric_diagram}    
\eeq
$T_{\mathrm{CY}_{m+n+3}}$ is a lattice polytope in $\mathbb{Z}^{m+n+2}$. In this lattice, the $T_{\mathrm{CY}_{m+2}}$ gets embedded in a hyperplane spanned by the first $m+1$ coordinates, while $T_{\mathrm{CY}_{n+2}}$ gets embedded in a hyperplane spanned by the last $n+1$ coordinates. These two hyperplanes are orthogonal and meet at a single point $(u_{0},v_{0})$. In other words, the final toric diagram $T_{\mathrm{CY}_{m+n+3}}$ is the convex hull of the set of points obtained by ``interlacing" $T_{\mathrm{CY}_{m+2}}$ and $T_{\mathrm{CY}_{n+2}}$ at the point $(u_0,v_0)$. \fref{examples_product_toric_diagrams} shows two examples of this construction. Higher dimensional examples are straightforward although, obviously, difficult to visualize. 

At first sight, the use of the term ``product" to refer to the operation that acts on the geometry as described above, might be slightly confusing. The resulting geometry is {\it not} the product of the two parent CYs. In particular, its dimension is not equal to the sum of the dimensions of the starting CYs. However, we feel that the term captures various aspects of the process and its sufficiently simple to justify its adoption. 

It is clear that the product of CYs can very easily produce quiver theories for extremely complicated geometries. Moreover, iterating the process, it becomes straightforward to deal with high dimensional geometries. We will present explicit examples in \sref{section_examples}.

There is substantial freedom in this construction. Given a desired CY$_{m+n+3}$, it can generally be decomposed into other CY$_{m+2}$ and CY$_{n+2}$ geometries in multiple ways (even with different values of $m$ and $n)$, there is a choice of toric phase for each of the parent geometries and of perfect matchings for the points $u_0$ and $v_0$. Therefore, generically, the CY product method can generate a large number of quiver theories for a given CY$_{m+n+3}$, reflecting the rich space of theories related by the corresponding order $(m+n+2)$ dualities.

\section{Product of Toric Calabi-Yaus: the Periodic Quiver}

\label{section_periodic_quiver} 

Having discussed the connection between the parent and product geometries, we now explain how to construct the periodic quiver for the product. The periodic quiver contains all the information defining the quiver theory, namely not only the quiver but also the superpotential. Having said that, in \sref{section_superpotential} we will present explicit rules for constructing the superpotential directly, without having to read it from the periodic quiver.

The starting point of the construction is the initial data discussed in the previous section. As already mentioned, choosing different toric phases for the two parent geometries and/or using different perfect matchings for the $u_0$ and $v_0$ points can result in different phases for the same product geometry. Similar freedom has been observed in other constructions such as $3d$ printing \cite{Franco:2018qsc} and it is natural to expect such different phases to be related by duality.  

As discussed in \sref{section_toric_quivers}, in order to simplify the product construction, given a perfect matching it is convenient to pick the polarization of the quiver in which  the perfect matching turns out to simply consist of the conjugates of all the fields in the quiver. We will do so here. Using the polarization of $P$ given by $p$ and the polarization of $Q$ given by $q$, we will define a polarization of the periodic quiver for $P_{p}\times Q_{p}$. As we will see later, this polarization in fact corresponds to a perfect matching of the product theory and corresponds to the point $(u_{0},v_{0})$. 
 
The periodic quiver of the product theory $P_{p}\times Q_{p}$ can be elegantly defined in terms of the action of the product operation on the basic elements of the parent quivers: nodes and fields. Below, we will use the following convention to denote nodes and fields in the different quivers: $i$ and $X$ for $P$, $j$ and $Y$ for $Q$ and $(i,j)$ and $Z$ for $P_{q}\times Q_{q}$. We have three possible products:

\paragraph{Node $\times$ node.}

\begin{figure}[H]
	\centering
	\includegraphics[width=4.5cm]{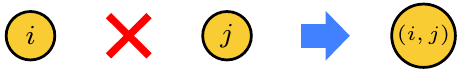}
\caption{Node $\times$ node.}
	\label{node_x_node}
\end{figure}

The product of nodes $i$ of $P$ and $j$ of $Q$ gives rise to a node $(i,j)$ of $P_{p}\times Q_{q}$. This process is illustrated in \fref{node_x_node}.

\paragraph{Field $\times$ node.}

\begin{figure}[H]
	\centering
	\includegraphics[width=15.5cm]{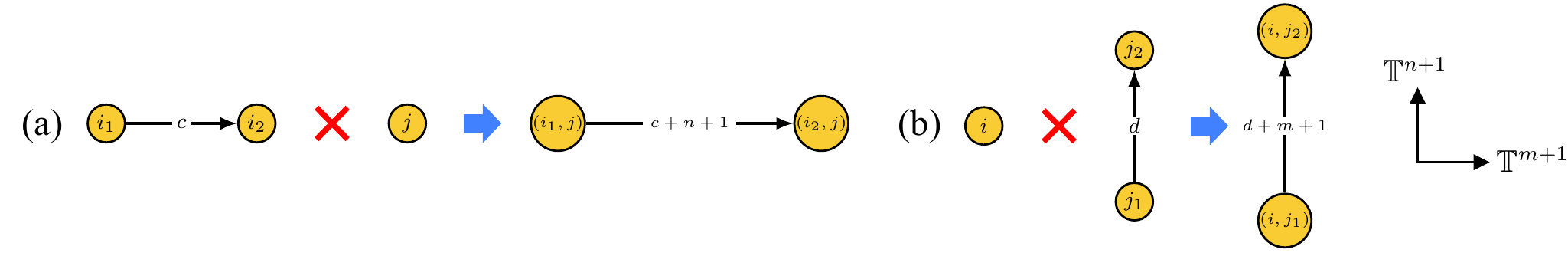}
\caption{Field $\times$ node.}
	\label{node_x_field}
\end{figure}

The product of a field $\bar{X}^{(c)}_{i_{1},i_{2}}$ of $P$ which is in $p$ with a node $j$ of $Q$ gives rise to a field $\bar{Z}^{(c+n+1)}_{(i_{1},j)(i_{2},j)}$ in $P_{p} \times Q_{q}$. Similarly, the product of a node $i$ of $P$ and a field $\bar{Y}^{(d)}_{j_{1}j_{2}}$ of $Q$ which is in $q$ gives rise to a field $\bar{Z}^{(d+m+1)}_{(i,j_{1})(i,j_{2})}$ in $P_{p}\times Q_{q}$.\footnote{For clarity, we have emphasized that we go over the fields $\bar{X}^{(c)}_{i_{1},i_{2}}$ of $P$ which are in $p$ and the fields $\bar{Y}^{(d)}_{j_{1}j_{2}}$ of $Q$ in $q$. However, given our choice of polarization determined by $p$ and $q$, these are simply the conjugates of {\it all} the fields in $P$ and $Q$.} \fref{node_x_field} represents this operation. The horizontal and vertical directions encode the $\mathbb{T}^{m+1}$ and $\mathbb{T}^{n+1}$ tori, respectively.

\paragraph{Field $\times$ field.}

\begin{figure}[H]
	\centering
	\includegraphics[width=10.2cm]{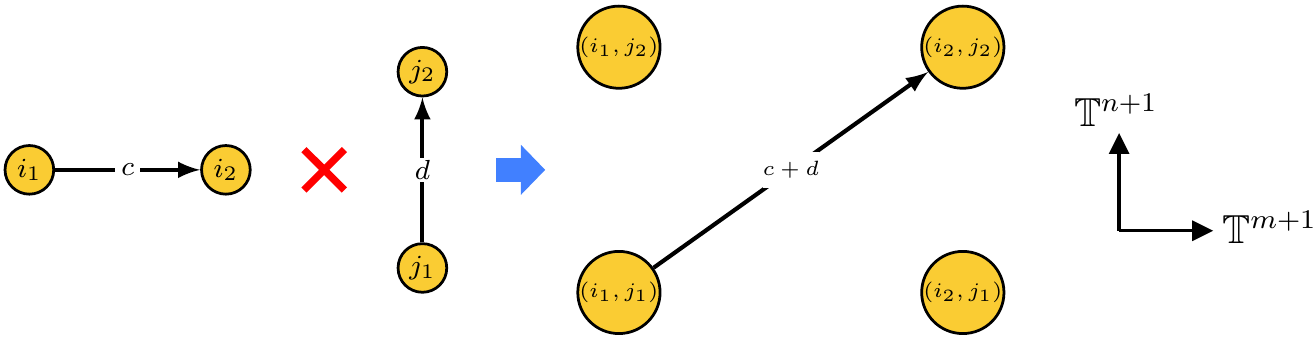}
\caption{Field $\times$ field.}
	\label{field_x_field}
\end{figure}

The product of a field $\bar{X}^{(c)}_{i_{1}i_{2}}$ of $P$ in $p$ with a field $\bar{Y}^{(d)}_{j_{1}j_{2}}$ of $Q$ in $q$ gives rise to a field $\bar{Z}_{(i_{1},j_{1})(i_{2},j_{2})}^{(c+d)}$.  \fref{field_x_field} represents this operation.

Table \ref{table_product} summarizes the product construction. This procedure not only generates the quiver for $P_{p} \times Q_{q}$ but also constructs its periodic quiver. This is because given an embedding of the periodic quiver $P$ in $\mathbb{T}^{m+1}$ and of $Q$ in $\mathbb{T}^{n+1}$, these rules result in an embedding of $P_{p} \times Q_{q}$ in $\mathbb{T}^{m+n+2}$.

 \begin{table}[ht]
        \setstretch{2}
        \centering
        \begin{tabular}{|c|c|c|}
            \hline
           \ \ \ \ \ $P$ \ \ \ \ \ & \ \ \ \ \ $Q$ \ \ \ \ \ & \ \ \ $P_{p} \times Q_{q}$ \ \ \ \\
            \hline
            $i$ & $j$ & $(i,j)$ \\ \hline
            $i$ & $\bar{Y}^{(d)}_{j_{1}j_{2}}$ & $\bar{Z}^{(d+m+1)}_{(i,j_{1})(i,j_{2})}$  \\
            $\bar{X}^{(c)}_{i_{1}i_{2}}$ & $j$ & $\bar{Z}^{(c+n+1)}_{(i_{1},j)(i_{2},j)}$ \\ \hline
            $\bar{X}^{(c)}_{i_{1}i_{2}}$ & $\bar{Y}^{(d)}_{j_{1}j_{2}}$ & $\bar{Z}_{(i_{1},j_{1})(i_{2},j_{2})}^{(c+d)}$ \\ \hline
        \end{tabular}
        \caption{Summary of the construction of the periodic quiver for $P_{p} \times Q_{q}$}
        \label{table_product}
        \end{table}

For the sake of completeness we also describe the conjugates of the fields we have written above. Their origin can be understood as follows:
         \begin{itemize}
             \item 
                The conjugate of $\bar{Z}_{(i,j_{1})(i,j_{2})}^{(d+m+1)}$ is $Z^{(n-d)}_{(i,j_{2})(i,j_{1})}$. It arises from the product between the node $i$ and the field $Y_{j_{2}j_{1}}^{(n-d)}$ which is not in $q$.
            \item
                The conjugate of $\bar{Z}^{(c+n+1)}_{(i_{1},j)(i_{2},j)}$ is $Z_{(i_{2},j)(i_{1},j)}^{(m-c)}$. It comes from the product between $X^{(m-c)}_{i_{2}i_{1}}$ which is not in $p$ and node $j$.
            \item
                The conjugate of $\bar{Z}_{(i_{1},j_{1})(i_{2},j_{2})}^{(c+d)}$ is $Z_{(i_{2},j_{2})(i_{1},j_{1})}^{(m+n+1-c-d)}$. It comes from the product between $X^{(m-c)}_{i_{2}i_{1}}$ and $Y^{(n-d)}_{j_{2}j_{1}}$.   
         \end{itemize}

It is important to note that at the end of this process there is no field that comes from the product of an $\bar{X}^{(c)}_{i_{1}i_{2}} \in p$ and a $Y_{j_{2}j_{1}}^{(d)} \notin q$ or vice versa. This makes the choice of $p$ and $q$ central to this construction.

\subsection{Anomaly Cancellation}

Let us begin checking the consistency of the CY product construction we have just introduced. In this section we will show that if $P$ and $Q$ satisfy the corresponding anomaly cancellation conditions, then so does $P_{p} \times Q_{q}$. We assume that the ranks of all nodes are equal to $N$ and normalize the anomaly by this number. We first enumerate all the fields that are charged under a given node $(i,j)$ of $P_{p} \times Q_{q}$ and consider their contributions to the anomaly. These fields are given by:
        \begin{enumerate}
            \item
                Product of incoming fields at $i$ in $P$ with node $j$ of $Q$. 
                \begin{enumerate}
                    \item 
                        If $\bar{X}^{(c)}_{i^{\prime}i}\in p$, then it gives rise to a field $\bar{Z}^{(c+n+1)}_{(i^{\prime},j)(i,j)}$ incoming at $(i,j)$ which contributes $(-1)^{c+n+1}$ to the anomaly.
                    \item
                         If $X_{i^{\prime}i} \notin p$, then it gives rise to a field $Z^{(c)}_{(i^{\prime},j)(i,j)}$ incoming at $(i,j)$ which contributes $(-1)^{c}$ to the anomaly.
                \end{enumerate}  
            \item
                Product of incoming field at $j$ in $Q$ with node $i$ of $P$. 
                \begin{enumerate}
                    \item 
                        If $\bar{Y}^{(d)}_{j^{\prime}j} \in q$, then it gives rise to a field $\bar{Z}^{(d+m+1)}_{(i,j^{\prime})(i,j)}$ incoming at $(i,j)$ which contributes $(-1)^{d+m+1}$ to the anomaly.
                    \item
                         If $Y_{j^{\prime}j} \notin q$, then it gives rise a field $Z^{(c)}_{(i,j^{\prime})(i,j)}$ incoming at $(i,j)$ which contributes $(-1)^{d}$ to the anomaly.
                \end{enumerate}  

\item Product of a field $\bar{X}^{(c)}_{i^{\prime} i }$ that is in $p$ with a field $\bar{Y}^{(d)}_{j^{\prime}j}$ that is in $q$. This gives rise to the incoming field $\bar{Z}_{(i^{\prime},j^{\prime})(i,j)}^{(c+d)}$ which contributes $(-1)^{c+d}$ to the anomaly. This is just the product of the the contribution to anomaly at $i$ of the incoming field $\bar{X}^{(c)}_{i^{\prime} i}$ and the contribution to the anomaly at $j$ of the incoming field $\bar{Y}^{(d)}_{j^{\prime}j}$. 
            \item
                Product of an outgoing field $\bar{X}^{(c)}_{i i^{\prime}}$ at $i$ that is in $p$ with an outgoing field $\bar{Y}^{(d)}_{jj^{\prime}}$ at $j$ that is in $q$. This gives rise to the outgoing field $\bar{Z}_{(i,j)(i^{\prime},j^{\prime})}^{(c+d)}$ at $(i,j)$. Its conjugate contributes $(-1)^{m+n+1-c-d}$ to the anomaly. This is {\it minus} the product of the contributions to the anomaly at $i$ of the incoming field $X^{(m-c)}_{i^{\prime}i}$ and the contribution to the anomaly at $j$ of the incoming field $Y^{(n-d)}_{j^{\prime}j}$.             
        \end{enumerate}
        Adding all these contributions, the anomaly at node $(i,j)$ becomes
\beq
            A = a_{\cancel{p}} + (-1)^{n+1}a_{p} + b_{\cancel{q}} + (-1)^{m+1}b_{q} + a_{p}b_{q}  - a_{\cancel{p}}b_{\cancel{q}} ~,
\eeq
where $a_{p}$ is the contribution to the anomaly by incoming fields at $i$ which are in $p$ and $a_{\cancel{p}}$ is the contribution to the anomaly by incoming fields that are not in $p$. Similarly, $b_{p}$ is the contribution to the anomaly at node $j$ by incoming fields that are in $q$, while $b_{\cancel{q}}$ is the contribution from the fields that are not in $q$.

At this point we distinguish three cases depending on the parity on $m$ and $n$.

\paragraph{Odd $m$ and $n$.}

In this case $A$ becomes
\beq
            A= a_{\cancel{p}} + a_{p} + b_{\cancel{q}} + b_{q} + a_{p}b_{q} - a_{\cancel{p}}b_{\cancel{q}} ~.
\eeq
        For odd $m$ and $n$, the anomaly cancellation conditions for $i$ in $P$ and $j$ in $Q$ respectively are
        \begin{align}
            a_{\cancel{p}} = -a_{p} && b_{\cancel{q}} = -b_{q}
        \end{align}
        Plugging these back into the expression for $A$ results in $A = 0$, which is the anomaly cancellation condition, since $m+n+1$ is odd.

\paragraph{Even $m$ and even $n$.} 

In this case $A$ becomes
\beq
            A_{net} = a_{\cancel{p}} - a_{p} + b_{\cancel{q}} - b_{q} + a_{p}b_{q} - a_{\cancel{p}}b_{\cancel{q}} ~.
\eeq
The anomaly cancellation conditions for $i$ and $j$ respectively are
            \begin{align}
                a_{\cancel{p}} = 2-a_{p} && b_{\cancel{q}} = 2-b_{q}
            \end{align}
            Plugging these back also results in $A = 0$, which is again the anomaly cancellation condition since $m+n+1$ is odd in this case, too.

\paragraph{Odd $m$ and even $n$.}

Lastly, in this case 
\beq
            A= a_{\cancel{p}} - a_{p} + b_{\cancel{q}} + b_{q} + a_{p}b_{q} - a_{\cancel{p}}b_{\cancel{q}} ~,
\eeq
The anomaly cancellation conditions at $i$ and $j$ are
        \begin{align}
            a_{\cancel{p}} = -a_{p} && b_{\cancel{q}} = 2-b_{q}
        \end{align}
        which gives $A = 2$, i.e. the anomaly cancellation condition is satisfied since $m+n+1$ is even for this case.

\section{Superpotential}
    
\label{section_superpotential}

The construction introduced in \sref{section_periodic_quiver}, produces the periodic quiver for $P_{p} \times Q_{q}$ from which, in principle, its superpotential can be read off. In general, this can be rather challenging. Therefore, in this section we introduce explicit rules for the direct construction of the superpotential.
 
The superpotential of the product theory takes the general form 
\beq
W = \mathcal{W}_P + \mathcal{W}_Q + \mathcal{W}_C + \mathcal{W}_{PQ} ~.      
\eeq
$\mathcal{W}_P$ and $\mathcal{W}_Q$ descend from the superpotentials of $P$ and $Q$, respectively. $\mathcal{W}_C$ consists of new cubic interactions. Finally, $\mathcal{W}_{PQ}$ depends on superpotentials of both $P$ and $Q$. We now describe each of them in detail.

\paragraph{$\mathcal{W}_P$: terms descending from the superpotential of $P$.} 

Let us consider a single term $T_{P}$ in the superpotential $W_P$ of the parent theory $P$. It has the general form
\beq
T_{P} = X^{(c_{1})}_{i_{1}i_{2}}X^{(c_{2})}_{i_{2}i_{3}}\cdots X^{(c_{k-1})}_{i_{k-1}i_{k}}\bar{X}^{(c_{k})}_{i_{k}i_{1}} ~,
\label{pterm_product_potential}    
\eeq
where $\sum_{n}c_{n} = m-1$ due to degree constraint. Our convention for the polarization makes the perfect matching $p$ manifest. The fields in $p$ appear as a single conjugated field per term in $W_P$. Furthermore, we will order the fields in every term such that the fields in $p$ occur last.

Every term $T_P$ gives rise to various terms in $\mathcal{W}_P$, as we now discuss. First, some of these terms correspond to the product between the fields in this term and a node $j$ of $Q$. They take the form
\beq
\sum_{j\in J}Z^{(c_{1})}_{(i_{1},j)(i_{2},j)}Z^{(c_{2})}_{(i_{2},j)(i_{3},j)}\cdots Z^{(c_{k-1})}_{(i_{k-1},j)(i_{k},j)}\bar{Z}^{(c_{k}+n+1)}_{(i_{k},j)(i_{1},j)} ~,
\label{potential_w_i_p_product}
\eeq
where the sum is over the set $J$ of nodes $j$ of $Q$. After this operation, the degree of the superpotential changes by $n+1$ and becomes $m+n$, as required for the superpotential of an $(m+n+1)$-graded quiver.

The additional terms descending from $T_P$ are constructed as follows. We first pick a field $X^{(c)}_{i^{\prime}i}$ from those in $T_{P}$. Since this field does not appear conjugated, it is obviously not contained in $p$. We also pick a field $Y_{j^{\prime}j}^{(d)}$ that is not in $q$. We then replace $X^{(c)}_{i^{\prime}i}$ in $T_{P}$ by its product with $Y_{j^{\prime}j}^{(d)}$, i.e. by $Z^{(c+d+1)}_{(i^{\prime},j^{\prime})(i,j)}$. This operation increases the degree by $d+1$. We also replace $\bar{X}^{(c_{k})}_{i_{k}i_{1}}$ by its product with $\bar{Y}_{j j^{\prime}}^{(n-d)}$, i.e. by $\bar{Z}^{(c_{k} + n-d)}_{(i_{k},j)(i_{1},j^{\prime})}$. This changes the degree by $n-d$. Finally, we simply replace the remaining fields in $T_{P}$ by their product with appropriate node in $Q$, which does not change the degrees since these fields are not in $p$. When combined, all these replacements change the degree of the superpotential term by $n+1$, as desired. Explicitly these terms are
        \begin{align}
            \sum_{\bar{Y}_{j j^{\prime}}^{(n-d)} \in q}\bigg[&Z^{(c_{1}+d+1)}_{(i_{1},j^{\prime})(i_{2},j)}Z^{(c_{2})}_{(i_{2},j)(i_{3},j)}Z^{(c_{3})}_{(i_{3},j)(i_{4},j)}\cdots Z^{(c_{k-1})}_{(i_{k-1},j)(i_{k},j)}\bar{Z}^{(c_{k}+n-d)}_{(i_{k},j)(i_{1},j^{\prime})}    \nonumber\\
            &\,\, + (-1)^{c_{1}}Z^{(c_{1})}_{(i_{1},j^{\prime})(i_{2},j^{\prime})}Z^{(c_{2}+d+1)}_{(i_{2},j^{\prime})(i_{3},j)}Z^{(c_{3})}_{(i_{3},j)(i_{4},j)}\cdots Z^{(c_{k-1})}_{(i_{k-1},j)(i_{k},j)}\bar{Z}^{(c_{k}+n-d)}_{(i_{k},j)(i_{1},j^{\prime})} +\cdots \nonumber \\
            &\,\,\,+ (-1)^{c_{1}+\cdots +c_{k-2}}Z^{(c_{1})}_{(i_{1},j^{\prime})(i_{2},j^{\prime})}Z^{(c_{2})}_{(i_{2},j^{\prime})(i_{3},j^{\prime})}Z^{(c_{3})}_{(i_{3},j^{\prime})(i_{4},j^{\prime})}\cdots Z^{(c_{k-1}+d+1)}_{(i_{k-1},j^{\prime})(i_{k},j)}\bar{Z}^{(c_{k}+n-d)}_{(i_{k},j)(i_{1},j^{\prime})}\bigg] 
            \label{potential_w_y_p_product}
        \end{align} 
        To obtain $\mathcal{W}_P$, we repeat this process for all the terms in $W_P$. In addition to the signs written above, we must include the signs with which the parent superpotential terms enter $W_P$.

\paragraph{$\mathcal{W}_Q$: terms descending from the superpotential in $Q$.}   
 
These terms are determined by the same procedure, after the exchange $(P,p) \leftrightarrow (Q,q)$. Let us present the final result. Every term $T_{Q}$ in the superpotential $W_Q$ of $Q$ is of the form:
\beq
        T_{Q} = Y^{(d_{1})}_{j_{1}j_{2}}Y^{(d_{2})}_{j_{2}j_{3}}\cdots Y^{(d_{k-1})}_{j_{k-1}j_{k}}\bar{Y}^{(d_{k})}_{j_{k}j_{1}} ~.
        \label{qterm_product_potential}    
\eeq  
As before, $T_{Q}$ gives rise to superpotential terms of two types, analogous to \eref{potential_w_i_p_product} and \eref{potential_w_y_p_product}. The first set of terms is
\beq
        \sum_{i\in I}Z^{(d_{1})}_{(i,j_{1})(i,j_{2})}Z^{(d_{2})}_{(i,j_{2})(i,j_{3})}\cdots Z^{(d_{l-1})}_{(i,j_{l-1})(i,j_{k})}\bar{Z}^{(d_{l}+m+1)}_{(i,j_{l})(i,j_{1})} ~,
        \label{potential_w_j_q_product}
\eeq
with $I$ the set of nodes of $P$.

The second set of terms is   
\begin{align}
        \sum_{\bar{X}_{i i^{\prime}}^{(m-c)} \in p}\bigg[&Z^{(c+d_{1}+1)}_{(i^{\prime},j_{1})(i,j_{2})}Z^{(d_{2})}_{(i,j_{2})(i,j_{3})}Z^{(d_{3})}_{(i,j_{3})(i,j_{4})}\cdots Z^{(d_{l-1})}_{(i,j_{l-1})(i,j_{l})}\bar{Z}^{(m-c + d_{k})}_{(i,j_{l})(i^{\prime},j_{1})}    \nonumber\\
        &\,\, + (-1)^{d_{1}}Z^{(d_{1})}_{(i^{\prime},j_{1})(i^{\prime},j_{2})}Z^{(c+d_{2}+1)}_{(i^{\prime},j_{2})(i,j_{3})}Z^{(d_{3})}_{(i,j_{3})(i,j_{4})}\cdots Z^{(d_{l-1})}_{(i,j_{l-1})(i,j_{l})}\bar{Z}^{(m-c +d_{l})}_{(i,j_{l})(i^{\prime},j_{1})}+\cdots+ \nonumber \\
        &\,\,\, +(-1)^{d_{1}+\cdots+d_{l-2}}Z^{(d_{1})}_{(i^{\prime},j_{1})(i^{\prime},j_{2})}Z^{(d_{2})}_{(i^{\prime},j_{2})(i^{\prime},j_{3})}Z^{(d_{3})}_{(i^{\prime},j_{3})(i^{\prime},j_{4})}\cdots Z^{(m-c+d_{l-1})}_{(i^{\prime},j_{l-1})(i,j_{l})}\bar{Z}^{(m-c +d_{l})}_{(i,j_{l})(i^{\prime},j_{1})}\bigg] \label{potential_w_x_q_product}
    \end{align}
Repeating this process for all the terms in $W_P$ , we obtain $\mathcal{W}_P$. Once again, we need to include the signs of the parent terms in $W_P$.

\paragraph{$\mathcal{W}_C$: new cubic interactions.}

This part of the superpotential consists of new cubic interactions. For every pair of fields $\bar{X}^{(c)}_{i_1 i_2}\in p$ and $\bar{Y}^{(d)}_{j_1 j_2}\in q$ we have a pair of cubic terms
\beq
(-1)^{c+d}\left[Z^{(n-d)}_{(i_2,j_2)(i_2,j_1)}Z^{(m-c)}_{(i_2,j_1)(i_1,j_2)}\bar{Z}^{(c+d)}_{(i_1,j_1)(i_2,j_2)} -Z^{(m-c)}_{(i_2,j_2)(i_1,j_2)}Z^{(n-d)}_{(i_1,j_2)(i_1,j_1)}\bar{Z}^{(c+d)}_{(i_1,j_1)(i_2,j_2)}\right] 
\label{cubic_terms}
\eeq
where the fields involved are descendants of $\bar{X}^{(c)}_{ii^{\prime}}$ and $\bar{Y}^{(d)}_{jj^{\prime}}$ via the rules in Table \ref{table_product}, or their conjugates. Namely,
\beq
Z^{(m-c)}_{(i_2,j_1)(i_1,j_1)} = X^{(m-c)}_{i_2 i_1} \times j_1 \ \ , \ \ Z^{(n-d)}_{(i_1,j_2)(i_1,j_1)} = i_1 \times  Y^{(n-d)}_{j_2 j_1} \ \ , \ \  \bar{Z}^{(c+d)}_{(i_1,j_1)(i_2,j_2)} = \bar{X}^{(c)}_{i_1 i_2} \times \bar{Y}^{(d)}_{j_1 j_2} ~.
\eeq
$\mathcal{W}_C$ is the sum of \eref{cubic_terms} over all the pairs of $\bar{X}^{(c)}_{i_1 i_2}$ and $\bar{Y}^{(d)}_{j_1 j_2}$.

\paragraph{$\mathcal{W}_{PQ}$: mixed terms.}
    \label{mixed_terms}
The last part of the superpotential involves contributions coming from $P$ and $Q$. A term $T_{P}$ in the superpotential of $P$ and a term $T_{Q}$ in the superpotential of $Q$ give rise to a number of terms in the superpotential of the product theory. $\mathcal{W}_{PQ}$ is the sum of all such terms. To describe them, let us first consider the special case in which both $T_{P}$ and $T_{Q}$ are cubic terms, i.e.
\beq
        T_{P} = X_{i_{1}i_{2}}^{(c_{1})}X_{i_{2}i_{3}}^{(c_{2})}\bar{X}^{(m-1-c_{1}-c_{2})}_{i_{3}i_{1}} \ \ \ , \ \ \ T_{Q} = Y_{j_{1}j_{2}}^{(d_{1})}Y_{j_{2}j_{3}}^{(d_{2})}\bar{Y}_{j_{3}j_{1}}^{(n-1-d_{1}-d_{2})} ~.  
\label{cubic_TP_and_TQ}
\eeq
In this case, they give rise to a single term that involves the pairwise product of fields,\footnote{It is useful to reflect on why we obtain a single term. First of all, we defined the polarizations of the parent theories such that every term in their superpotentials contains a single conjugated field. In addition, following the rules introduced in \sref{section_periodic_quiver}, we cannot multiply unbarred and barred fields. As a result, there are not multiple possibilities associated to cyclic permutations of the fields in \eref{cubic_TP_and_TQ}.} i.e.
\beq
        (-1)^{m+n+c_{2}+d_{2}}Z_{(i_{1},j_{1}),(i_{2},j_{2})}^{(c_{1}+d_{1}+1)}Z^{(c_{2}+d_{2}+1)}_{(i_{2},j_{2}),(i_{3},j_{3})}\bar{Z}^{(n+m-2-c_{1}-c_{2}-d_{1}-d_{2})}_{(i_{3},j_{3}),(i_{1},j_{1})} ~.
\eeq

If $T_{P}$ and/or $T_{Q}$ are of order greater than 3, no such simple terms can be written. The reason is that the pairwise product of fields is only possible if they have the same order and the resulting terms will have correct degree, i.e. $m+n$, if and only if $T_{P}$ and $T_{Q}$ are cubic.\footnote{It is interesting to compare this to the B-model computation of the superpotential: cubic terms are special in that they correspond to $m_{2}$ of the $A_{\infty}$ algebra, which is composition of maps, while higher order terms correspond to higher $m_{k}$, which are more involved.}

One way of addressing this issue is to turn $T_{P}$ and $T_{Q}$ into a sum of cubic terms and mass terms, by integrating in auxiliary massive fields. Then we can construct $\mathcal{W}_{PQ}$ as described above, consisting exclusively of terms descending from the cubic terms. The final quiver and superpotential can then be obtained by integrating out the massive fields. 

Naively, it might seem that this procedure dramatically changes our construction. A massive field in $P$ gives rise to one descendant for every field or node of $Q$ and vice versa. Nevertheless, it can be verified that all these descendants are massive, resulting in the same quiver we would have obtained without integrating in massive fields. Therefore, we can use the rule for cubic terms above as the starting point to efficiently compute the rules for higher order terms. The result is that there are $\binom{k-1}{2}\binom{l-1}{2}$ terms in $\mathcal{W}_{PQ}$ descending from terms $T_{P}$ of order $k$ and $T_{Q}$ of order $l$. All these terms are of order $k+l-3$. We provide a thorough discussion of these terms and the first few steps of this iteration in Appendix \ref{mixed_potential_terms_appendix}.

\paragraph{The geometry of the product theory.}

It is relatively straight forward, yet quite laborious, to show that the desired geometry \eref{product_toric_diagram} arises as the classical moduli space of the $P_{p} \times Q_{q}$ theory we have constructed.\footnote{The notion of moduli space has been extended to general $m$ in \cite{Franco:2017lpa}.} We present the proof in Appendix \ref{section_products_geometry}.

\subsection{Kontsevich Bracket}

As another consistency check of our construction, let us verify that the superpotential we have written satisfies $\{W,W\} = 0$,  where
\beq
        \{W,W\} = 2\sum_{\bar{Z}^{(b)}_{(i,j)(i^{\prime},j^{\prime})}} \pdv{W}{Z^{(b)}_{(i^{\prime},j^{\prime})(i,j)}}\pdv{W}{\bar{Z}^{(m+n+1-b)}_{(i,j)(i^{\prime},j^{\prime})}} ~.
\eeq
To do this, we divide $\{W,W\}$ into eight pieces, 
\beq
             \{W,W\} = 2(KB_{P} + KB_{Q} + KB_{PC} + KB_{QC} + KB_{PQ} + KB_{PQP} + KB_{PQQ} + KB_{PQC}) ~,
\eeq
each of which vanishes individually.

$KB_{P} = \frac{1}{2}\{\mathcal{W}_P,\mathcal{W}_P\}$ is the contribution that arises exclusively due to $\mathcal{W}_P$. Explicitly, its nontrivial terms are
\beq
             KB_{P} = \sum_{j\in J}\sum_{\bar{X}^{(c)}_{i_{1}i_{2}} \in p}\pdv{\mathcal{W}_P}{Z^{(m-c)}_{(i_{2},j)(i_{1},j)}}\pdv{\mathcal{W}_P}{\bar{Z}_{(i_{1},j)(i_{2,j})}^{(c+n+1)}} + \sum_{\bar{Y}_{j_{1}j_{2}}^{(d)}\in q}\sum_{\bar{X}^{(c)}_{i_{1}i_{2}} \in p}\pdv{\mathcal{W}_P}{Z^{(m+n+1-c-d)}_{(i_{2},j_{2})(i_{1},j_{1})}}\pdv{\mathcal{W}_P}{\bar{Z}_{(i_{1},j_{1})(i_{2},j_{2})}^{(c+d)}} ~.
\eeq
It is straightforward to show that $KB_{P}$ vanishes if the superpotential $W_P$ of $P$ satisfies $\{W_P,W_P\} = 0$. The reason is that the terms in $KB_{P}$ descend from the terms of $\{W_P,W_P\}$ in a manner that is analogous to how terms in $\mathcal{W}_P$ descend from terms in $W_P$ and the signs in \eref{potential_w_y_p_product} are such that the required cancellations still occur.

        Similarly, $KB_{Q} = \frac{1}{2}\{\mathcal{W}_Q,\mathcal{W}_Q\}$ is
\beq
            KB_{Q} = \sum_{i\in I}\sum_{\bar{Y}^{(d)}_{j_{1}j_{2}} \in q}\pdv{\mathcal{W}_Q}{Z^{(n-d)}_{(i,j_{2})(i,j_{1})}}\pdv{\mathcal{W}_Q}{\bar{Z}_{(i,j_{1})(i,j_{2})}^{(d+m+1)}} + \sum_{\bar{X}_{i_{1}i_{2}}^{(c)}\in p}\sum_{\bar{Y}^{(d)}_{j_{1}j_{2}} \in q}\pdv{\mathcal{W}_Q}{Z^{(m+n+1-c-d)}_{(i_{2},j_{2})(i_{1},j_{1})}}\pdv{\mathcal{W}_Q}{\bar{Z}_{(i_{1},j_{1})(i_{2},j_{2})}^{(c+d)}}~,
\eeq
        and it vanishes if the superpotential $W_Q$ of $Q$ satisfies $\{W_Q,W_Q\} = 0$. 

$KB_{PC}$ and $KB_{QC}$ involve the Kontsevich bracket between $\mathcal{W}_P$ and $\mathcal{W}_Q$ with $\mathcal{W}_C$. Explicitly, $KB_{PC} = \frac{1}{2}(\{\mathcal{W}_P,\mathcal{W}_C\}+ \{\mathcal{W}_C,\mathcal{W}_P\}) $ and $KB_{QC} =\frac{1}{2}( \{\mathcal{W}_Q,\mathcal{W}_C\}+ \{\mathcal{W}_C,\mathcal{W}_Q\})$. They reduce to
        \begin{align}
            KB_{PC} & =  & \sum_{j\in J}\sum_{\bar{X}^{(c)}_{i_{1}i_{2}} \in p}\pdv{\mathcal{W}_C}{Z^{(m-c)}_{(i_{2},j)(i_{1},j)}}\pdv{\mathcal{W}_P}{\bar{Z}_{(i_{1},j)(i_{2,j})}^{(c+n+1)}} + \sum_{\bar{Y}_{j_{1}j_{2}}^{(d)}\in q}\sum_{\bar{X}^{(c)}_{i_{1}i_{2}} \in p}\pdv{\mathcal{W}_P}{Z^{(m+n+1-c-d)}_{(i_{2},j_{2})(i_{1},j_{1})}}\pdv{\mathcal{W}_C}{\bar{Z}_{(i_{1},j_{1})(i_{2},j_{2})}^{(c+d)}}\nonumber \\
            KB_{QC} & = & \sum_{i \in I}\sum_{\bar{Y}^{(d)}_{j_{1}j_{2}} \in q}\pdv{\mathcal{W}_C}{Z^{(n-d)}_{(i,j_{2})(i,j_{1})}}\pdv{\mathcal{W}_Q}{\bar{Z}_{(i,j_{1})(i,j_{2})}^{(d+m+1)}} + \sum_{\bar{X}_{i_{1}i_{2}}^{(c)}\in p}\sum_{\bar{Y}^{(d)}_{j_{1}j_{2}} \in q}\pdv{\mathcal{W}_Q}{Z^{(m+n+1-c-d)}_{(i_{2},j_{2})(i_{1},j_{1})}}\pdv{\mathcal{W}_C}{\bar{Z}_{(i_{1},j_{1})(i_{2},j_{2})}^{(c+d)}}
        \end{align} 
Both $KB_{PC}$ and $KB_{QC}$ vanish independently of any conditions on $W_P$ and $W_Q$. This can be verified directly using the explicit form of $\mathcal{W}_C$.

Let us now consider $KB_{PQ} = \frac{1}{2}\{\mathcal{W}_{PQ},\mathcal{W}_{PQ}\}$. Its non-trivial part is
\beq
KB_{PQ} = \sum_{\bar{X}^{(c)}_{i_{1}i_{2}}\in p}\sum_{\bar{Y}^{(d)}_{j_{1}j_{2}} \in q}\pdv{\mathcal{W}_{PQ}}{Z^{(m+n+1-c-d)}_{(i_{2},j_{2})(i_{1},j_{1})}}\pdv{\mathcal{W}_{PQ}}{\bar{Z}_{(i_{1},j_{1})(i_{2},j_{2})}^{(c+d)}}~.
\eeq
First, let us consider the case in which $W_P$ and $W_Q$ are cubic, since in this case $\mathcal{W}_{PQ}$ comes just from the pairwise product of fields, as explained earlier. In this case, both $\{W_P,W_P\}$ and $\{W_Q,W_Q\}$ are entirely quartic and a term in $KB_{PQ}$ comes from the pairwise product of fields from a term in $\{W_P,W_P\}$ and a term in $\{W_Q,W_Q\}$, As result, $KB_{PQ}$ vanishes.

To show that $KB_{PQ}$ vanishes even when $W_P$ and $W_Q$ are not cubic, we can rewrite  $W_P$ and $W_Q$ as sums of cubic terms and mass terms by appropriately integrating in massive fields and using the argument above. There is an added subtlety: after integrating in these massive fields, $\{W_P,W_P\}$ and $\{W_Q,W_Q\}$ vanish only after using the equations of motion for massive fields. This is enough for our purposes, and it can be shown that $KB_{PQ}$ vanishes once we integrate out massive fields from the product theory.

All the remaining contributions, $KB_{PQP},KB_{PQQ}$ and $KB_{PQC}$, involve $\mathcal{W}_{PQ}$ and therefore it is convenient to express $W_P$ and $W_Q$ as a sum of cubic terms and mass terms. Explicitly, they are 
        \begin{align}
             KB_{PQP} &= \frac{1}{2}\left(\{\mathcal{W}_{PQ},\mathcal{W}_P\} + \{\mathcal{W}_P,\mathcal{W}_{PQ}\}\right) \nonumber\\
             KB_{QPQ} &= \frac{1}{2}\left(\{\mathcal{W}_{PQ},\mathcal{W}_Q\} + \{\mathcal{W}_Q,\mathcal{W}_{PQ}\}\right) \nonumber \\
             KB_{PQC} &= \frac{1}{2}(\{\mathcal{W}_P,\mathcal{W}_Q\} + \{\mathcal{W}_Q,\mathcal{W}_P\} + \{\mathcal{W}_{PQ},\mathcal{W}_C\} + \{\mathcal{W}_C,\mathcal{W}_{PQ}\})
        \end{align} 
        A lengthy but straightforward bookkeeping calculation shows that all of these contributions vanish up to the equations of motion for massive fields. $KB_{PQP}$ vanishes as a result of $\{W_P,W_P\}=0$, while vanishing of $KB_{PQQ}$ follows from $\{W_Q,W_Q\}=0$. Lastly, $KB_{PQC}$ vanishes independently of any restriction on $W_P$ and $W_Q$.

\subsection{Toric Condition} 

To conclude our discussion of the superpotential, we now show that our construction is such that if $P$ and $Q$ satisfy the toric condition then $P_{q} \times Q_{q}$ also does so. We do so by considering the different ways a field of degree $m+n$ can arise in the superpotential of $P_{p} \times Q_{q}$. It is useful to note that all such terms must come from $\mathcal{W}_P,\mathcal{W}_Q$ and $\mathcal{W}_C$, but not from $\mathcal{W}_{PQ}$. As explained in Appendix \ref{mixed_potential_terms_appendix}, every term in $\mathcal{W}_{PQ}$ contains two fields coming from the product of a field not in $p$ and a field not in $q$. The degrees of such fields are greater or equal to 1, so none of these terms can contain a degree $m+n$ field. The different scenarios are:
\begin{itemize}
\item A field of degree $m-1$, $\bar{X}^{(m-1)}_{i_{1}i_{2}}\in p$. Its product with a node $j$ of $Q$ gives rise to a field $\bar{Z}^{(m+n)}_{(i_{1},j)(i_{2},j)}$ of degree $m+n$. This field only appears in $\mathcal{W}_P$, in the form shown in \eref{potential_w_i_p_product}. Therefore, if $\bar{X}^{(m-1)}_{i_{1}i_{2}}$ participates in two terms with opposite signs, then so does $\bar{Z}^{(m+n)}_{(i_{1},j)(i_{2},j)}$. Similarly, if there is a field $\bar{Y}_{j_{1}j_{2}}^{(n-1)}\in q$, its product with a node $i$ of $P$ gives rise to $\bar{Z}^{(m+n)}_{(i,j_{1})(i,j_{2})}$. It only participates in $\mathcal{W}_P$, as shown in \eref{potential_w_j_q_product}, namely in two terms with opposite sign.

\item The product of a conjugate chiral $\bar{X}^{(m)}_{i_{1}i_{2}}\in p$ and a conjugate chiral field $\bar{Y}^{(n)}_{j_{1}j_{2}}\in q$ gives rise to a field $\bar{Z}^{(m+n)}_{(i_{1},j_{1})(i_{2},j_{2})}$ of degree $m+n$. Since conjugate chiral fields do not appear in the superpotential, $\bar{Z}^{(m+n)}_{(i_{1},j_{1})(i_{2},j_{2})}$ does not appear in $\mathcal{W}_P$ or $\mathcal{W}_Q$. It only appears in two terms of $\mathcal{W}_C$ with opposite sign as shown in \eref{cubic_terms}.

\item The product of a field $X^{(m-1)}_{i_{1}i_{2}}\notin p$ and a conjugate chiral field $Y_{j_{1}j_{2}}^{(n)}\notin q$ gives a field $Z^{(m+n)}_{(i_{1},j_{1})(i_{2},j_{2})}$ of degree $m+n$. Since $X^{(m-1)}_{i_{1}i_{2}}$ appears in two terms with opposite sings in $W_P$, $Z^{(m+n)}_{(i_{1},j_{1})(i_{2},j_{2})}$ appears in two terms of the final superpotential with opposite signs. These terms arise as described by \eref{potential_w_y_p_product}. Since $Y_{j_{1}j_{2}}^{(n)}$ is a conjugate chiral, it does not appear in $W_Q$, which implies that $Z^{(m+n)}_{(i_{1},j_{1})(i_{2},j_{2})}$ does not appear in $\mathcal{W}_Q$. It does not appear in $\mathcal{W}_C$, either.

Similarly, the product of a conjugate chiral field $X^{(m)}_{i_{1}i_{2}}\notin p$ and $Y_{j_{1}j_{2}}^{(n-1)}\notin q$ gives rise to $Z^{(m+n)}_{(i_{1},j_{1})(i_{2},j_{2})}$, which only appears in two terms with opposite signs. These terms are in $\mathcal{W}_Q$, specifically among those described in \eref{potential_w_x_q_product}.   
\end{itemize}

The discussion above covers all the fields of degree $m+n$. We conclude that the product between an $m$-graded toric phase $P$ and an $n$-graded toric phase $Q$ using arbitrary perfect matchings is an $(m+n+1)$-graded toric phase.

\section{Examples}

\label{section_examples}

In this section we illustrate the product construction with two explicit examples. The first theory we will construct is the well-known phase 2 of $F_{0}$ \cite{Feng:2002zw}.\footnote{By phase 2, we mean the phase whose quiver is shown in \eref{F0_2_quiver}. Various papers label the two phases of $F_0$ in different ways.} The second example is a product of the conifold quiver theory with itself, which results in a  $0d$ $\mathcal{N}=1$ matrix model. While, to our knowledge, this the first time the second theory appears in the literature, our primary goal is to demonstrate the simplicity of this procedure.

\subsection{$F_{0}$}
\label{subsec_F0}
        
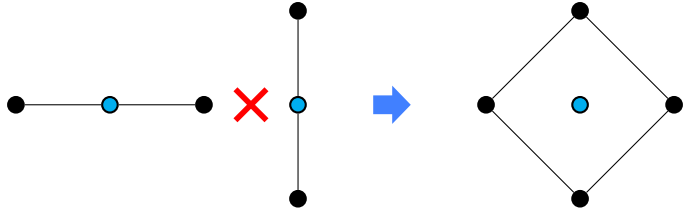
\begin{figure}[H]
                 \centering
                \begin{tikzpicture}[scale=1.25]
                    \tikzstyle{every node}=[circle,thick,fill=black,draw,minimum size = 0.5em , inner sep=1pt,font=\scriptsize]
                    \draw (0,0) node[fill = cyan](a){};
                    \draw (1,0) node(b){};
                    \draw (-1,0) node(e){};
                    \draw [black](e) -- (a) -- (b);
                    \draw (1.5,0) node[cross=8pt,red, line width = 0.75mm]{};
                    \draw (2,0) node[fill = cyan](a){};
                    \draw (2,1) node(b){};
                    \draw (2,-1) node(e){};
                    \draw [black](e) -- (a) -- (b);
                    \draw [blue2 , -{Triangle[angle = 90:3.5mm]},line width = 2.5mm] (2.8,0)--(3.2,0);
                    \draw (5,0) node[fill = cyan](a){};
                    \draw (6,0) node(b){};
                    \draw (4,0) node(c){};
                    \draw (5,1) node(d){};
                    \draw (5,-1) node(e){};
                    \draw (b)--(d)--(c)--(e)--(b);
                \end{tikzpicture}  
                \caption{The toric diagram of $F_{0}$ can be obtained as the product of two copies of the toric diagram of $\mathbb{C}^{2}/\mathbb{Z}_{2}$. In both cases we use the central point of the toric diagram to take the product.}
                \label{F_0_product_example}
             \end{figure}

Let us consider the complex cone over $F_{0}$ CY 3-fold, or $F_0$ for short. The $m=1$, i.e. $4d$ $\mathcal{N}=1$, quiver theories for this geometry have been extensively studied in the literature (see e.g. \cite{Feng:2002zw}). The toric diagram for $F_0$ can be constructed as the product of two copies of $\mathbb{C}^{2}/\mathbb{Z}_{2}$ using  one of the two perfect matchings for the central point in each case, as illustrated in \fref{F_0_product_example}.

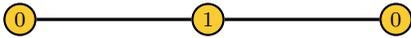
\begin{figure}[H]
                 \centering
                \begin{tikzpicture}[scale=2.5,decoration={markings,mark=at position 0.75 with {\arrow{latex}}}]
                    \tikzstyle{every node}=[circle,thick,fill=yellow2,draw,minimum size = 1em , inner sep=1pt,font=\scriptsize]
                    \draw (0,0) node(a){$0$};
                    \draw (1,0) node(b){$1$};
                    \draw (2,0) node(c){$0$};
                    \draw [ very thick ](a) -- (b);
                    \draw [ very thick ](b) -- (c);
                \end{tikzpicture}
                \caption{The periodic quiver for $\mathbb{C}^{2}/\mathbb{Z}_{2}$}
                \label{c2_z2_periodic_quiver}
        \end{figure}

The $m=0$, i.e. $6d$ $\mathcal{N}=(1,0)$, quiver theory of the parent $\mathbb{C}^{2}/\mathbb{Z}_{2}$ geometry consists of two $U(N)$ gauge groups with two hypermultiplets stretching between them, as shown in \fref{c2_z2_periodic_quiver}. 

This theory has 4 perfect matchings, which translate into the 4 ways in which we can orient the 2 hypermultiplets. Two of them correspond to the two endpoints of the toric diagram (shown on the left of \fref{F_0_product_example}) while the other 2 correspond to the central point. As a result, we have 2 perfect matching choices for the central point of each of the $\mathbb{C}^{2}/\mathbb{Z}_{2}$ factors. But the 2 central perfect matchings are conjugates of each other and as a result any choice of perfect matchings gives the same theory up to chiral conjugation.\footnote{We note that $m=0$ is the only case for which the conjugates of the field in a perfect matching also form a perfect matching. This is only possible because there is no superpotential.}  

\begin{figure}[H]
                 \centering
                \begin{tikzpicture}[scale = 1.5]
                    \tikzstyle{every node}=[circle,thick,fill=yellow2,draw,inner sep=1pt,font=\scriptsize]
                    \draw (0,0) node(a){$0$};
                    \draw (1,0) node(b){$1$};
                    \draw (2,0) node(c){$0$};
                    \draw [very thick,-latex] (a) -- (b);
                    \draw [very thick,-latex] (c) -- (b);
                    \draw (2.5,0) node[cross=6pt,red,ultra thick]{};
                    \draw (3,-1) node(a){$0$};
                    \draw (3,0) node(b){$1$};
                    \draw (3,1) node(c){$0$};
                    \draw [very thick,-latex] (a) -- (b);
                    \draw [very thick,-latex] (c) -- (b);
                    \draw [blue2 , -{Triangle[angle = 90:3.5mm]},line width = 2.5mm] (3.8,0)--(4.2,0);
                    \draw (5,-1) node(a){$\scalebox{0.5}{(0,0)}$};
                    \draw (5,0) node(b){$\scalebox{0.5}{(0,1)}$};
                    \draw (5,1) node(c){$\scalebox{0.5}{(0,0)}$};
                    \draw (6,-1) node(d){$\scalebox{0.5}{(1,0)}$};
                    \draw (6,0) node(e){$\scalebox{0.5}{(1,1)}$};
                    \draw (6,1) node(f){$\scalebox{0.5}{(1,0)}$};
                    \draw (7,-1) node(g){$\scalebox{0.5}{(0,0)}$};
                    \draw (7,0) node(h){$\scalebox{0.5}{(0,1)}$};
                    \draw (7,1) node(i){$\scalebox{0.5}{(0,0)}$};
                    \draw [blue2 , -{Triangle[angle = 90:3.5mm]},line width = 2.5mm] (6,-1.8)--(6,-2.2);
                    \begin{scope}[shift={(0,-4)}]
                    \draw (5,-1) node(a){$\scalebox{0.5}{(0,0)}$};
                    \draw (5,0) node(b){$\scalebox{0.5}{(0,1)}$};
                    \draw (5,1) node(c){$\scalebox{0.5}{(0,0)}$};
                    \draw (6,-1) node(d){$\scalebox{0.5}{(1,0)}$};
                    \draw (6,0) node(e){$\scalebox{0.5}{(1,1)}$};
                    \draw (6,1) node(f){$\scalebox{0.5}{(1,0)}$};
                    \draw (7,-1) node(g){$\scalebox{0.5}{(0,0)}$};
                    \draw (7,0) node(h){$\scalebox{0.5}{(0,1)}$};
                    \draw (7,1) node(i){$\scalebox{0.5}{(0,0)}$};
                    \draw [very thick,-latex] (a) -- (d);
                    \draw [very thick,-latex] (g) -- (d);
                    \draw [very thick,-latex] (b) -- (e);
                    \draw [very thick,-latex] (h) -- (e);
                    \draw [very thick,-latex] (c) -- (f);
                    \draw [very thick,-latex] (i) -- (f); 
                    \draw [very thick,-latex] (a) -- (b);
                    \draw [very thick,-latex] (c) -- (b);
                    \draw [very thick,-latex] (d) -- (e);
                    \draw [very thick,-latex] (f) -- (e);
                    \draw [very thick,-latex] (g) -- (h);
                    \draw [very thick,-latex] (i) -- (h);
                    \end{scope} 
                    \draw [blue2 , -{Triangle[angle = 90:3.5mm]},line width = 2.5mm] (4.2,-4)--(3.8,-4);      
                    \begin{scope}[shift={(-4,-4)}]
                    \draw (5,-1) node(a){$\scalebox{0.5}{(0,0)}$};
                    \draw (5,0) node(b){$\scalebox{0.5}{(0,1)}$};
                    \draw (5,1) node(c){$\scalebox{0.5}{(0,0)}$};
                    \draw (6,-1) node(d){$\scalebox{0.5}{(1,0)}$};
                    \draw (6,0) node(e){$\scalebox{0.5}{(1,1)}$};
                    \draw (6,1) node(f){$\scalebox{0.5}{(1,0)}$};
                    \draw (7,-1) node(g){$\scalebox{0.5}{(0,0)}$};
                    \draw (7,0) node(h){$\scalebox{0.5}{(0,1)}$};
                    \draw (7,1) node(i){$\scalebox{0.5}{(0,0)}$};
                    \draw [very thick,-latex] (a) -- (d);
                    \draw [very thick,-latex] (g) -- (d);
                    \draw [very thick,-latex] (b) -- (e);
                    \draw [very thick,-latex] (h) -- (e);
                    \draw [very thick,-latex] (c) -- (f);
                    \draw [very thick,-latex] (i) -- (f); 
                    \draw [very thick,-latex] (a) -- (b);
                    \draw [very thick,-latex] (c) -- (b);
                    \draw [very thick,-latex] (d) -- (e);
                    \draw [very thick,-latex] (f) -- (e);
                    \draw [very thick,-latex] (g) -- (h);
                    \draw [very thick,-latex] (i) -- (h);
                    \draw [very thick,-latex] (e) -- (a);
                    \draw [very thick,-latex] (e) -- (c);
                    \draw [very thick,-latex] (e) -- (g);
                    \draw [very thick,-latex] (e) -- (i);
                    \end{scope}
                \end{tikzpicture}
                \caption{A product of periodic quivers resulting in phase 2 of $F_{0}$. }
                \label{F_0_product_quiver}
             \end{figure}
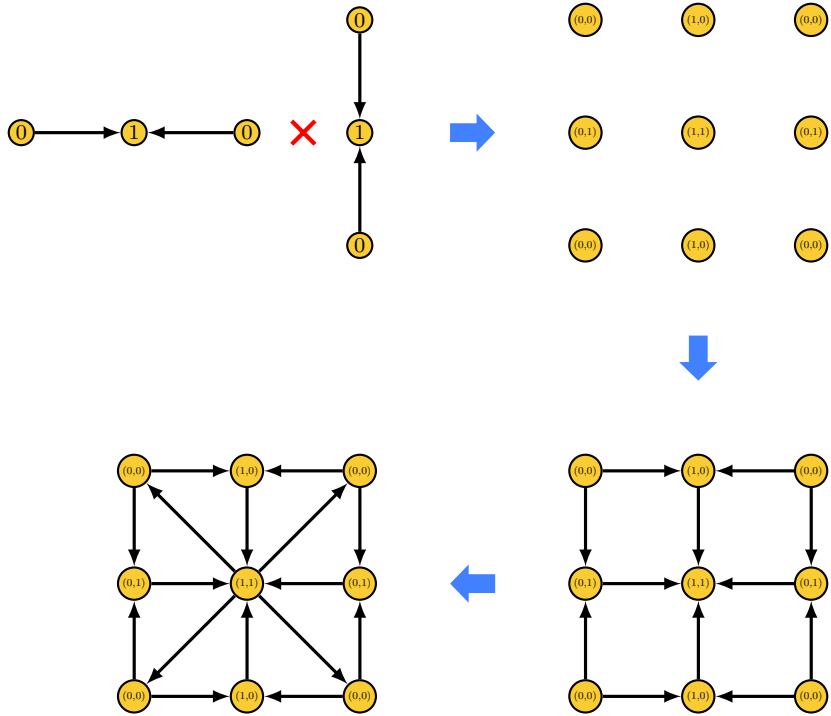

The product of the periodic quivers is presented in \fref{F_0_product_quiver}. The first step shows the two parent $6d$ $\mathcal{N}=(1,0)$ quivers. The arrows are oriented to indicate the choice of perfect matchings. The second step shows the nodes of $F_{0}$ that arise from the product of nodes in the parent theories. In the third step, we add vertical fields (which come from the product of a node in the first parent and a field in the second one) and horizontal fields (which come from the product of a field in the first parent and a node in the second one). The last step adds the diagonal fields that arise from the product of a field in the first parent with a field in the second one.

The result is the phase 2 of $F_{0}$ \cite{Feng:2002zw}. Since in this the parent theories do not have superpotentials, the final superpotential only consists of the new cubic terms that arise in the product. These terms can be straightforwardly read from the minimal plaquettes of the quiver. 
   
\begin{figure}[H]
             \centering
            \begin{tikzpicture}[scale=2.5,decoration={markings,mark=at position 0.75 with {\arrow{latex}}}]
                \tikzstyle{every node}=[circle,thick,fill=yellow2,draw,minimum size = 0.5em , inner sep=1pt,font=\scriptsize]
                \draw (0,0) node(a){$\scalebox{0.75}{(0,0)}$};
                \draw (1,0) node(b){$\scalebox{0.75}{(1,0)}$};
                \draw (0,1) node(c){$\scalebox{0.75}{(0,1)}$};
                \draw (1,1) node(d){$\scalebox{0.75}{(1,1)}$};
                \draw [very thick, postaction = {decorate}](a) -- node[rectangle , minimum size = 0.6em , pos = 0.25, draw = none , fill = white , inner sep = 0, font = \tiny]{$2$} (b);
                \draw [very thick, postaction = {decorate}](a) -- node[rectangle , minimum size = 0.6em , pos = 0.25, draw = none , fill = white , inner sep = 0, font = \tiny]{$2$} (c);
                \draw [very thick, postaction = {decorate}](b) -- node[rectangle , minimum size = 0.6em , pos = 0.25, draw = none , fill = white , inner sep = 0, font = \tiny]{$2$} (d);
                \draw [very thick, postaction = {decorate}](c) -- node[rectangle , minimum size = 0.6em , pos = 0.25, draw = none , fill = white , inner sep = 0, font = \tiny]{$2$} (d);
                \draw [very thick, postaction = {decorate}](d) -- node[rectangle , minimum size = 0.6em , pos = 0.25, draw = none , fill = white , inner sep = 0, font = \tiny]{$4$} (a);
            \end{tikzpicture}
            \caption{The quiver for phase 2 of $F_{0}$}
            \label{F0_2_quiver}
    \end{figure}
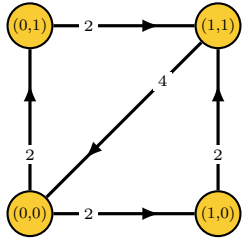

For completeness, in \fref{F0_2_quiver} we show the standard quiver for this theory. Its superpotential is 

\begin{align}
        W &= X^{+}_{(0,0)(0,1)}X^{+}_{(0,1)(1,1)}X^{--}_{(1,1)(0,0)} - X^{+}_{(0,0)(1,0)}X^{+}_{(1,0)(1,1)}X^{--}_{(1,1)(0,0)} \nonumber \\
        & + X^{+}_{(0,0)(1,0)}X^{-}_{(1,0)(1,1)}X^{-+}_{(1,1)(0,0)} - X^{-}_{(0,0)(0,1)}X^{+}_{(0,1)(1,1)}X^{-+}_{(1,1)(0,0)} \nonumber \\
        & + X^{-}_{(0,0)(1,0)}X^{+}_{(1,0)(1,1)}X^{+-}_{(1,1)(0,0)} - X^{+}_{(0,0)(0,1)}X^{-}_{(0,1)(1,1)}X^{+-}_{(1,1)(0,0)}  \nonumber \\
        & + X^{-}_{(0,0)(0,1)}X^{-}_{(0,1)(1,1)}X^{++}_{(1,1)(0,0)} - X^{-}_{(0,0)(1,0)}X^{-}_{(1,0)(1,1)}X^{++}_{(1,1)(0,0)}
    \end{align}
    
\paragraph{An Infinite Family: $F^{(m)}_{0}$.}
 The process discussed above can be continued inductively to get an infinite family of toric CY $(m+2)$-folds indexed by $m$. The toric diagram for $F^{(m)}_{0}$ is
            \begin{align}
                \begin{array}{c}
                (0,\ldots,0) \\
                (\pm 1,0,\ldots, 0) \\
                \vdots \\
                (0,\ldots,0,\pm1)
                 \end{array}
            \end{align}
 This family was first introduced in \cite{Closset:2018axq}, where the corresponding quiver theories were also constructed.           
            
 Roughly speaking the periodic quiver for $F_{0}^{(m)}$ corresponds to 
            \begin{align}
                  \left(\mathord{\begin{tikzpicture}[baseline = -0.65ex,scale = 2]
                    \tikzstyle{every node}=[circle,thick,fill=yellow2,draw,inner sep=1.25pt,font=\scriptsize]
                    \draw (0,0) node(a){0};
                    \draw (1,0) node(b){1};
                    \draw (2,0) node(c){0};
                    \draw [very thick,-latex] (a) -- (b);
                    \draw [very thick,-latex] (c) -- (b);
                \end{tikzpicture}}\right)^{m+1}
             \end{align} 
This is of course not a complete description except for $m=1$ because, at every step, to construct a periodic quiver for $F^{(n)}_{0}$ we need to choose a perfect matching for $F^{(n-1)}_{0}$. This freedom hints at the existence of multiple phases of $F_{0}^{(m)}$ for $m > 1$ and it is natural to expect that different choices of perfect matching lead to different phases related by the dualities discussed in \sref{subsec_dualities}.\footnote{For example, $F_0^{(2)}$ is also known as $Q^{1,1,1}/\mathbb{Z}_2$. This theory has 14 toric phases, which were classified in \cite{Franco:2018qsc}.}

The quiver theory $Q^{(m)}$ of one particular phase of $F^{(m)}_{0}$ can be constructed inductively as follows
             \begin{align}
                 Q^{(0)} = \mathord{\begin{tikzpicture}[baseline = -0.65ex,scale = 1.75]
                    \tikzstyle{every node}=[circle,thick,fill=yellow2,draw,inner sep=1.25pt,font=\scriptsize]
                    \draw (0,0) node(a){0};
                    \draw (1,0) node(b){1};
                    \draw (2,0) node(c){0};
                    \draw [very thick] (a) -- (b);
                    \draw [very thick] (c) -- (b);
                \end{tikzpicture}} & \ \ \ \ &p^{(0)} = \mathord{\begin{tikzpicture}[baseline = -0.65ex,scale = 1.75]
                    \tikzstyle{every node}=[circle,thick,fill=yellow2,draw,inner sep=1.25pt,font=\scriptsize]
                    \draw (0,0) node(a){0};
                    \draw (1,0) node(b){1};
                    \draw (2,0) node(c){0};
                    \draw [very thick,-latex] (a) -- (b);
                    \draw [very thick,-latex] (c) -- (b);
                \end{tikzpicture}}\nonumber \\
               Q^{(m+1)} = Q^{(m)}_{p^{(m)}} \times Q^{(0)}_{p^{(0)}} & & p^{(m+1)} = p^{(m)} \times p^{(0)}  \label{fom_product_construction}
             \end{align}
where we use the product perfect matching $p \times q$ of $P_{p}\times Q_{q}$ as defined in Appendix \ref{section_products_geometry}. This phase of $F_{0}^{(m)}$ was discussed at length in \cite{Closset:2018axq,Franco:2019bmx}, to which we refer for details.

\subsection{Conifold $\times$ Conifold}

\begin{figure}[ht]
	\centering
	\includegraphics[width=10cm]{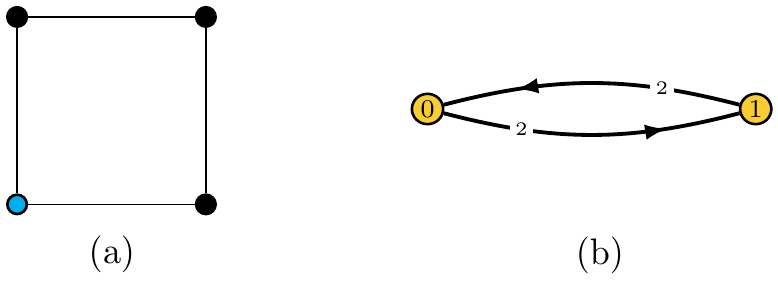}
\caption{a) Toric diagram and b) quiver for the conifold.}
	\label{toric_diagram_quiver_conifold}
\end{figure}

The conifold is one of the most thoroughly studied toric CY 3-folds. Its toric diagram is shown in \fref{toric_diagram_quiver_conifold}. The corresponding gauge theory was constructed in the seminal work \cite{Klebanov:1998hh}. It consists of two $U(N)$ gauge groups and four bifundamental chiral fields $X_{01}$, $\tilde{X}_{01}$, $X_{10}$, and $\tilde{X}_{10}$, as shown in \fref{toric_diagram_quiver_conifold}. The superpotential is
\beq
W_{con} = X_{01}X_{10}\tilde{X}_{01}\tilde{X}_{10} - \tilde{X}_{01}X_{10}X_{01}\tilde{X}_{10} \label{W_conifold}
\eeq

\begin{figure}[H]
                 \centering
                \begin{tikzpicture}[scale = 4,decoration={markings,mark=at position 0.75 with {\arrow{latex}}}]
                    \tikzstyle{every node}=[circle,thick,fill=yellow2,draw,inner sep=2pt,font=\small]
                    \draw (0,1) node(a){$0$};
                    \draw (1,1) node(b){$1$};
                    \draw (1,0) node(c){$2$};
                    \draw (0,0) node(d){$3$};
                    \draw [very thick, postaction = {decorate} , bend left = 15](a) to node[rectangle , minimum size = 0.6em , pos = 0.25, draw = none , fill = white , inner sep = 0, font = \tiny]{$2$}(b);
                    \draw [very thick,postaction = {decorate}, bend left = 15] (b) to (a);
                    \draw [very thick, postaction = {decorate} , bend left = 15](c) to node[rectangle , minimum size = 0.6em , pos = 0.25, draw = none , fill = white , inner sep = 0, font = \tiny]{$2$}(d);
                    \draw [very thick,postaction = {decorate}, bend left = 15] (d) to (c);
                    \draw [very thick, postaction = {decorate} , bend left = 10](a) to node[rectangle , minimum size = 0.6em , pos = 0.25, draw = none , fill = white , inner sep = 0, font = \tiny]{$2$}(c);
                    \draw [very thick,postaction = {decorate}, bend left = 10] (c) to (a);
                    \draw [very thick, postaction = {decorate} , bend left = 10](b) to node[rectangle , minimum size = 0.6em , pos = 0.25, draw = none , fill = white , inner sep = 0, font = \tiny]{$2$}(d);
                    \draw [very thick,postaction = {decorate}, bend left = 10] (d) to (b);
                    \draw [very thick,postaction = {decorate}] (d) to (a);
                    \draw [very thick,postaction = {decorate},red] (b) to (a);
                    \draw [very thick,postaction = {decorate},red] (d) to (c);
                    \draw [very thick,postaction = {decorate},red] (c) to (a);
                    \draw [very thick,postaction = {decorate},red] (d) to (b);
                    \draw [very thick, postaction = {decorate} , red , bend left = 15](d) to node[rectangle , minimum size = 0.6em , pos = 0.25, draw = none , fill = white , inner sep = 0, font = \tiny]{$4$}(a);
                    \draw [very thick, postaction = {decorate} , red , bend left = 15](a) to node[rectangle , minimum size = 0.6em , pos = 0.25, draw = none , fill = white , inner sep = 0, font = \tiny]{$5$}(d);
                     \draw [very thick, postaction = {decorate} , red , bend left = 15](b) to node[rectangle , minimum size = 0.6em , pos = 0.25, draw = none , fill = white , inner sep = 0, font = \tiny]{$3$}(c);
                    \draw [very thick, postaction = {decorate} , red , bend left = 15](c) to node[rectangle , minimum size = 0.6em , pos = 0.25, draw = none , fill = white , inner sep = 0, font = \tiny]{$3$}(b);
                \end{tikzpicture}
                \caption{Quiver for the conifold$\times$conifold. Black arrows have degree $0$ and red arrows have degree $2.$ They correspond to $0d$ $\mathcal{N}=1$ chiral and Fermi fields, respectively.}
                \label{conifold_conifold_quiver}
            \end{figure}
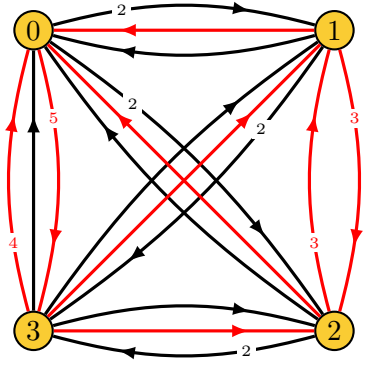

This theory has $4$ perfect matchings, each of them consists of one of the chiral fields and corresponds to a corner of the toric diagram. Given the symmetry between the perfect matchings, the result is independent of which perfect matching we use for the product, up to relabeling. We will therefore drop the reference to the perfect matching and refer to this theory as conifold$\times$conifold. Without loss of generality, we choose the toric diagrams of the two conifolds to coincide at the origin. The conifold$\times$conifold is therefore a toric CY 5-fold with toric diagram
            \begin{align}
                & (0,0,0,0) && (1,0,0,0) && (0,1,0,0) && (1,1,0,0)\nonumber \\ &(0,0,0,1) \nonumber \\ & (0,0,1,0) \nonumber\\ & (0,0,1,1)      
            \end{align}
where we have indicated the two conifold factors as the row and column. Table \ref{conifold_conifold_field_table} summarizes the nodes and fields in the product $0d$ $\mathcal{N}=1$ matrix model.\footnote{See e.g. \cite{Franco:2016tcm} for the basics of $0d$ $\mathcal{N}=1$ gauge theories.} The corresponding quiver is shown in \fref{conifold_conifold_quiver}.
               
            \begin{table}[ht]
                \setstretch{2}
                \centering
                \begin{tabular}{|c|cccccc|}
                \hline
                    & 0 & 1 & $X_{01}^{(0)}$ & $\tilde{X}_{01}^{(0)}$ & $X_{10}^{(0)}$ & $\bar{\tilde{X}}_{01}^{(1)}$ \\
  \hline
                  0 & $(0,0) \sim 0$ & $(0,1) \sim 1$ &  $Z_{01}^{(0)}$ & $\tilde{Z}_{01}^{(0)}$ & $Z_{10}^{(0)}$ & $\bar{\Lambda}_{01} ^{(1)}$\\  
                  1 & $(1,0) \sim 2$ & $(1,1) \sim 3$ & $Z_{23}^{(0)}$ & $\tilde{Z}_{23}^{(0)}$ & $Z_{32}^{(0)}$ & $\bar{\Lambda}_{23}^{(1)}$ \\
                  $X_{01}^{(0)}$ & $Z_{02}^{(0)}$  & $Z_{13}^{(0)}$ & $\bar{\Lambda}_{03} ^{(1)}$ & $\bar{\Sigma}_{03} ^{(1)}$ &$\bar{\Lambda}_{12}^{(1)}$ & $\Sigma_{03}^{(2)}$\\
                    $\tilde{X}_{01}^{(0)}$ & $\tilde{Z}_{02}^{(0)}$ & $\tilde{Z}_{13}^{(0)}$ & $\bar{\Gamma}_{03} ^{(1)}$ & $\bar{\Delta}_{03} ^{(1)}$ & $\bar{\Gamma}_{12}^{(1)}$ & $\Delta_{03}^{(2)}$ \\
                    $X_{10}^{(0)}$ & $Z_{20}^{(0)}$ & $Z_{31}^{(0)}$ & $\bar{\Lambda}_{21}^{(1)}$ & $\bar{\Sigma}_{21}^{(1)}$ & $\bar{\Lambda}_{30}^{(1)}$ & $\Sigma_{21}^{(2)}$\\
                     $\bar{\tilde{X}}_{01}^{(1)}$ & $\bar{\Lambda}_{02}^{(1)}$ & $\bar{\Lambda}_{13}^{(1)}$ & $\Gamma_{03} ^{(2)}$ & $\Omega_{03}^{(2)}$ & $\Gamma_{12}^{(2)}$ & $\bar{Z}_{03}^{(3)}$\\
                \hline
                \end{tabular}
                \caption{Summary of how the nodes and fields in the conifold$\times$conifold theory descend from the two parents. For simplicity, we converted the pairs of indices labeling nodes in the product to single indices. We also indicate the degree of the fields as a superindex. We use Latin and Greek letters to indicate chiral and Fermi fields, respectively.}
                \label{conifold_conifold_field_table}
            \end{table}

\paragraph{Superpotential.} 
Since the periodic quiver in this case lives on $\mathbb{T}^{4}$ we cannot display it diagrammatically. Instead, we can construct the superpotential explicitly using prescription given in \sref{section_superpotential}. We divide the total superpotential into four parts 
\beq
W = \mathcal{W}_{1} + \mathcal{W}{2} + \mathcal{W}_C + \mathcal{W}_{12} ~,
\eeq
where $\mathcal{W}_{1}$ and $\mathcal{W}_{2}$ come from the first and second conifold factors respectively, $\mathcal{W}_C$ contains the new cubic terms and $\mathcal{W}_{12}$ contains the mixed terms. Recall that in Table \ref{conifold_conifold_field_table} we used Latin and Greek letters to indicate chiral and Fermi fields, respectively. With this in mind the various parts of superpotential are:  

\medskip

\paragraph{\underline{$\mathcal{W}_{1}$}:} 

Since there are only two terms in the superpotential of the conifold we write the descendants of each of them separately. We thus write $\mathcal{W}_{1} = \mathcal{W}_{1+} - \mathcal{W}_{1-}$, with $\mathcal{W}_{1+}$ and $\mathcal{W}_{1-}$ the descendants of the positive and negative terms, respectively. We get
              \begin{align}
                  \mathcal{W}_{1+} &= Z_{01}Z_{10}\tilde{Z}_{01}\Lambda_{10} + Z_{23}Z_{32}\tilde{Z}_{23}\Lambda_{32} +\bar{\Lambda}_{03}Z_{32}\tilde{Z}_{23}\bar{\Sigma}_{30} + Z_{01}\bar{\Lambda}_{12}\tilde{Z}_{23}\bar{\Sigma}_{30} + Z_{01}Z_{10}\bar{\Sigma}_{03}\Sigma_{30} \nonumber \\
                        & + \bar{\Gamma}_{03}Z_{32}\tilde{Z}_{23}\bar{\Delta}_{30} + Z_{01}\bar{\Lambda}_{12}\tilde{Z}_{23}\bar{\Delta}_{30} + Z_{01}Z_{10}\bar{\Delta}_{03}\bar{\Delta}_{30} +  \bar{\Lambda}_{21}Z_{10}\tilde{Z}_{01}\bar{\Sigma}_{12} + Z_{23}\bar{\Lambda}_{30}\tilde{Z}_{01}\bar{\Sigma}_{12} \nonumber \\
                        & + Z_{23}Z_{32}\bar{\Sigma}_{21}\bar{\Sigma}_{12} + \Gamma_{03}Z_{32}\tilde{Z}_{23}Z_{30} + Z_{01}\Gamma_{12}\tilde{Z}_{23}Z_{30} + Z_{01}Z_{10}\Omega_{03}Z_{30}
              \end{align}
and
              \begin{align}
                  \mathcal{W}_{1-} &= \tilde{Z}_{01}Z_{10}Z_{01}\Lambda_{10} + \tilde{Z}_{23}Z_{32}Z_{23}\Lambda_{32} + \bar{\Sigma}_{03}Z_{32}Z_{23}\bar{\Sigma}_{30} + \tilde{Z}_{01}\bar{\Lambda}_{12}\tilde{Z}_{23}\bar{\Sigma}_{30} + \tilde{Z}_{01}Z_{10}\bar{\Lambda}_{03}\bar{\Sigma}_{30}\nonumber \\
                  & + \bar{\Delta}_{03}Z_{32}Z_{23}\bar{\Delta}_{30} + \tilde{Z}_{01}\bar{\Gamma}_{12}Z_{23}\bar{\Delta}_{30} + \tilde{Z}_{01}Z_{10}\bar{\Gamma}_{03}\bar{\Delta}_{30} + \bar{\Lambda}_{21}Z_{10}Z_{01}\bar{\Sigma}_{12} + \tilde{Z}_{23}\bar{\Lambda}_{30}Z_{01}\bar{\Sigma}_{12} \nonumber\\
                  & + \tilde{Z}_{23}Z_{32}\bar{\Lambda}_{21}\bar{\Sigma}_{12} + \Omega_{03}Z_{32}Z_{23}Z_{30} + \tilde{Z}_{01}\Gamma_{12}Z_{23}Z_{30} + \tilde{Z}_{01}Z_{10}\Gamma_{03}Z_{30}
              \end{align}
              
\medskip           
              
\paragraph{\underline{$\mathcal{W}_{2}$}:} 

Similarly, $\mathcal{W}_{2} = \mathcal{W}_{2+} - \mathcal{W}_{2-}$, with the two parts being
              \begin{align}
                   \mathcal{W}_{2+} &= Z_{02}Z_{20}\tilde{Z}_{02}\Lambda_{20} + Z_{13}Z_{31}\tilde{Z}_{13}\Lambda_{31} + \bar{\Lambda}_{03}Z_{31}\tilde{Z}_{13}\bar{\Gamma}_{30} + Z_{02}\bar{\Lambda}_{21}\tilde{Z}_{13}\bar{\Gamma}_{30} + Z_{02}Z_{20}\bar{\Gamma}_{03}\bar{\Gamma}_{30} \nonumber\\
                   & + \bar{\Sigma}_{03}Z_{31}\tilde{Z}_{13}\bar{\Omega}_{30} + Z_{02}\bar{\Sigma}_{21}\tilde{Z}_{13}\bar{\Omega}_{30} + Z_{01}Z_{10}\bar{\Delta}_{03}\bar{\Omega}_{30} + \bar{\Lambda}_{12}Z_{20}\tilde{Z}_{02}\bar{\Gamma}_{21} + Z_{13}\bar{\Lambda}_{30}\tilde{Z}_{02}\bar{\Gamma}_{21} \nonumber \\ 
                   & + Z_{13}Z_{31}\bar{\Gamma}_{12}\bar{\Gamma}_{21} + \Sigma_{30}Z_{31}\tilde{Z}_{13}Z_{30} + Z_{02}\Sigma_{21}\tilde{Z}_{13}Z_{30} + Z_{02}Z_{20}\Delta_{03}Z_{30} \nonumber\\[.15cm]
                   \mathcal{W}_{2-} &= \tilde{Z}_{02}Z_{20}Z_{02}\Lambda_{20} + \tilde{Z}_{13}Z_{31}Z_{13}\Lambda_{31} + \bar{\Gamma}_{03}Z_{31}Z_{13}\bar{\Gamma}_{30} + \tilde{Z}_{02}\bar{\Lambda}_{21}Z_{13}\bar{\Gamma}_{30} + \tilde{Z}_{02}Z_{20}\bar{\Lambda}_{03}\bar{\Gamma}_{30} \nonumber \\
                   & + \bar{\Sigma}_{03}Z_{31}Z_{13}\bar{\Omega}_{30} + \tilde{Z}_{02}\bar{\Sigma}_{21}Z_{13}\bar{\Omega}_{30} + \tilde{Z}_{02}Z_{20}\bar{\Sigma}_{03}\bar{\Omega}_{30} + \bar{\Gamma}_{12}Z_{20}Z_{02}\bar{\Gamma}_{21} + \tilde{Z}_{13}\bar{\Lambda}_{30}Z_{02}\bar{\Gamma}_{21} \nonumber\\
                   & + \tilde{Z}_{13}Z_{31}\bar{\Lambda}_{12}\bar{\Gamma}_{21} + \Delta_{03}Z_{31}Z_{13}Z_{30} + \tilde{Z}_{02}\Sigma_{21}Z_{13}Z_{30} + \tilde{Z}_{02}Z_{20}\Sigma_{03}Z_{30}
               \end{align}

\paragraph{\underline{$\mathcal{W}_C$}:}

As explained in \sref{section_superpotential} there are two cubic terms in the superpotential of $P_{p}\times Q_{q}$ for every pair of fields $\bar{X}^{(c)}_{i_{1},j_{1}} \in p$ and $\bar{Y}^{(d)}_{i_{2},j_{2}} \in Q$. In the present case, these terms are:
\begin{align}
                \renewcommand{\arraystretch}{1.25}
                \begin{array}{|c|c|c|c|c|}
                    \hline 
                           & X_{01} & \tilde{X}_{01} & X_{10} & \bar{\tilde{X}}_{01} \\
                    \hline
                    X_{01} & \phantom{-} Z_{01}Z_{13}\Lambda_{30} &  \phantom{-} Z_{01}\tilde{Z}_{13}\Gamma_{30} & \phantom{-} Z_{23}Z_{31}\Lambda_{12} &  \phantom{-} Z_{01}\bar{\Lambda}_{13}\bar{\Gamma}_{30}\\
                           &  - Z_{02}Z_{23}\Lambda_{30} & - \tilde{Z}_{02}Z_{23}\Gamma_{30} & - Z_{20}Z_{01}\Lambda_{12} & - \bar{\Lambda}_{02}Z_{23}\bar{\Gamma}_{30}  \\
                    \hline
                    \tilde{X}_{01} &\phantom{-} \tilde{Z}_{01}Z_{13}\Sigma_{30} &\phantom{-} \tilde{Z}_{01}\tilde{Z}_{13}\Delta_{30} &\phantom{-} \tilde{Z}_{23}Z_{31}\Sigma_{12} & \phantom{-} \tilde{Z}_{01}\bar{\Lambda}_{13}\bar{\Omega}_{30} \\
                        & - Z_{02}\tilde{Z}_{23}\Sigma_{30} & - \tilde{Z}_{02}\tilde{Z}_{23}\Delta_{30} & - Z_{20}\tilde{Z}_{01}\Sigma_{12} & - \bar{\Lambda}_{02}\tilde{Z}_{23}\bar{\Omega}_{30} \\
                    \hline
                    X_{10} & \phantom{-} Z_{10}Z_{02}\Lambda_{21}  & \phantom{-} Z_{10}\tilde{Z}_{02}\Gamma_{21}  & \phantom{-} Z_{32}Z_{20}\Lambda_{03} & \phantom{-} Z_{10}\bar{\Lambda}_{02}\bar{\Gamma}_{21}\\
                          & - Z_{13}Z_{32}\Lambda_{21} & - \tilde{Z}_{13}Z_{32}\Gamma_{21} & - Z_{31}Z_{10}\Lambda_{03} & - \bar{\Lambda}_{13}Z_{32}\bar{\Gamma}_{21} \\
                    \hline
                    \bar{\tilde{X}}_{01} & \phantom{-} \bar{\Lambda}_{01}Z_{13}\bar{\Sigma}_{30} & \phantom{-} \bar{\Lambda}_{01}\tilde{Z}_{13}\bar{\Delta}_{30} & \phantom{-} \bar{\Lambda}_{23}Z_{31}\bar{\Sigma}_{12}  & \phantom{-}\bar{\Lambda}_{01}\bar{\Lambda}_{13}Z_{30}  \\
                                       & - Z_{02}\bar{\Lambda}_{23}\bar{\Sigma}_{30} &  - \tilde{Z}_{02}\bar{\Lambda}_{23}\bar{\Delta}_{30} & - Z_{20}\bar{\Lambda}_{01}\bar{\Sigma}_{12} & - \bar{\Lambda}_{02}\bar{\Lambda}_{23}Z_{30}  \\
                    \hline 
                \end{array}
                \end{align} 
$\mathcal{W}_C$ is the sum of all these terms.

\paragraph{\underline{$W_{12}$}:} 

As explained in \sref{section_superpotential} and Appendix \ref{mixed_potential_terms_appendix}, for every pair of terms $T_{P}$ and $T_{Q}$, there are terms in the product superpotential that combine them. For every pair of quartic terms, there are $9$ quintic terms. As in the case of $\mathcal{W}_{1}$ and $\mathcal{W}_{2}$, we write the corresponding terms separately. So we decompose $\mathcal{W}_{12}$ as
\beq
\mathcal{W}_{12} = \mathcal{W}_{++} + \mathcal{W}_{+-} + \mathcal{W}_{-+} + \mathcal{W}_{--}~,
\eeq
where the signs correspond to the signs of the parent terms in the two conifolds. The individual contributions are: 
\begin{align}
        \mathcal{W}_{++} &= Z_{02}Z_{23}\bar{\Lambda}_{30}\bar{\Delta}_{03}Z_{30} - Z_{02}\bar{\Lambda}_{21}\bar{\Gamma}_{12}\tilde{Z}_{23}Z_{30} - Z_{02}\bar{\Lambda}_{21}Z_{10}\bar{\Delta}_{03}Z_{30} \nonumber\\ 
        &- Z_{01}\bar{\Lambda}_{12}\bar{\Sigma}_{21}\tilde{Z}_{13}Z_{30} + \bar{\Lambda}_{03}Z_{32}\bar{\Sigma}_{21}\tilde{Z}_{13}Z_{30} - Z_{01}\bar{\Lambda}_{12}Z_{20}\bar{\Delta}_{03}Z_{30} \nonumber\\
        & + \bar{\Lambda}_{03}Z_{32}Z_{20}\bar{\Delta}_{03}Z_{30} + \bar{\Lambda}_{03}\bar{\Lambda}_{30}\tilde{Z}_{02}\tilde{Z}_{23}Z_{30} + \bar{\Lambda}_{03}Z_{31}\bar{\Gamma}_{12}\tilde{Z}_{23}Z_{30} \nonumber\\[.15cm]
        \mathcal{W}_{+-} &= - Z_{02}\tilde{Z}_{23}\bar{\Lambda}_{30}\bar{\Gamma}_{03}Z_{30} + Z_{02}\bar{\Sigma}_{21}\bar{\Gamma}_{12}Z_{23}Z_{30} + Z_{02}\bar{\Sigma}_{21}Z_{10}\bar{\Gamma}_{03}Z_{30} \nonumber \\
        & + \tilde{Z}_{01}\bar{\Lambda}_{12}\bar{\Lambda}_{21}\tilde{Z}_{13}Z_{30} - \bar{\Sigma}_{03}Z_{32}\bar{\Lambda}_{21}\tilde{Z}_{13}Z_{30} + \tilde{Z}_{01}\bar{\Lambda}_{12}Z_{20}\bar{\Gamma}_{03}Z_{30} \nonumber \\
        & - \bar{\Sigma}_{03}Z_{32}Z_{20}\bar{\Gamma}_{03}Z_{30} - \bar{\Sigma}_{03}\bar{\Lambda}_{30}\tilde{Z}_{02}Z_{23}Z_{30} - \bar{\Sigma}_{03}Z_{31}\bar{\Gamma}_{12}Z_{23}Z_{30}\nonumber \\[.15cm]
        \mathcal{W}_{-+} &= - \tilde{Z}_{02}Z_{23}\bar{\Lambda}_{30}\bar{\Sigma}_{03}Z_{30} + \tilde{Z}_{02}\bar{\Lambda}_{21}\bar{\Lambda}_{12}\tilde{Z}_{23}Z_{30} + \tilde{Z}_{02}\bar{\Lambda}_{21}Z_{10}\bar{\Sigma}_{03}Z_{30} \nonumber\\
        & + Z_{01}\bar{\Gamma}_{12}\bar{\Sigma}_{21}Z_{13}Z_{30} - \bar{\Gamma}_{03}Z_{32}\bar{\Sigma}_{21}Z_{13}Z_{30} + Z_{01}\bar{\Gamma}_{12}Z_{20}\bar{\Sigma}_{03}Z_{30}  \nonumber \\
        & - \bar{\Gamma}_{03}Z_{32}Z_{20}\bar{\Sigma}_{03}Z_{30} - \bar{\Gamma}_{03}\bar{\Lambda}_{30}Z_{02}\tilde{Z}_{23}Z_{30} - \bar{\Gamma}_{03}Z_{31}\bar{\Lambda}_{12}\tilde{Z}_{23}Z_{30}  \nonumber \\[.15cm]         
         \mathcal{W}_{--} &= \tilde{Z}_{02}\tilde{Z}_{23}\bar{\Lambda}_{30}\bar{\Lambda}_{03}Z_{30} - \tilde{Z}_{02}\bar{\Sigma}_{21}\bar{\Lambda}_{12}Z_{23}Z_{30} - \tilde{Z}_{02}\bar{\Sigma}_{21}Z_{10}\bar{\Lambda}_{03}Z_{30} \nonumber\\
          & - \tilde{Z}_{01}\bar{\Gamma}_{12}\bar{\Lambda}_{21}Z_{13}Z_{30} + \bar{\Delta}_{03}Z_{32}\bar{\Lambda}_{21}Z_{13}Z_{30} - \tilde{Z}_{01}\bar{\Gamma}_{12}Z_{20}\bar{\Lambda}_{03}Z_{30} \nonumber \\
          & +\bar{\Delta}_{03}Z_{32}Z_{20}\bar{\Lambda}_{03}Z_{30} + \bar{\Delta}_{03}\bar{\Lambda}_{30}Z_{02}Z_{23}Z_{30} + \bar{\Delta}_{03}Z_{31}\bar{\Lambda}_{12}Z_{23}Z_{30}
    \end{align}
    This completes our description of the superpotential. All in all, it consists of $124$ terms. Of these, $38$ are $J$-terms, i.e. they contain precisely one degree $m-1$ field (namely degree 2 in this case) and the rest are chiral fields. Each one of the $19$ degree $m-1$ fields (see the quiver in \fref{conifold_conifold_quiver}) appear in two of these terms with opposite sign, so the superpotential satisfies the toric condition. Finally, with some effort we can verify that the Kontsevich bracket $\{W,W\}$ vanishes.

\section{Relation to Other Constructions}

\label{section_relation_to_other_constructions}

We now briefly discuss how the product construction relates to other known methods for determining the quiver theories corresponding to a given geometry.

\subsection{Algebraic Dimensional Reduction}

Algebraic dimensional reduction is an algorithm for constructing the quiver theory for $\mathrm{CY}_{m+2}\times \mathbb{C}$ starting from the quiver theory for $\mathrm{CY}_{m+2}$ \cite{Franco:2017lpa}. It generalizes dimensional reduction from $6d$ $\mathcal{N}=(1,0)$ theories to $4d$ $\mathcal{N}=2$ theories ($m=0\to m=1$), from $4d$ $\mathcal{N}=1$ theories to $2d$ $\mathcal{N}=(2,2)$ theories ($m=1\to m=2$) and from $2d$ $\mathcal{N}=(0,2)$ theories to $0d$ $\mathcal{N}=2$ theories ($m=2\to m=3$)\footnote{In all these cases, the dimensionally reduced theories have more than $2^{3-m}$ supercharges.} to arbitrary $m$.     

\begin{figure}[H]
                \centering
                \begin{tikzpicture}[scale=1]
                    \tikzstyle{every node}=[circle,thick,fill=yellow2,draw,inner sep=2pt,font=\small]
                    \draw (0,0) node(a){$0$};
                    \draw (3,0) node(b){$0$};
                    \draw[very thick](a) -- (b);
                    \draw (5,0) node(a){$0$};
                    \draw (8,0) node(b){$0$};
                    \draw[very thick,-latex](a) -- (b);
                    \draw (10,0) node(a){$0$};
                    \draw (13,0) node(b){$0$};
                    \draw[very thick,latex-](a) -- (b);
                \end{tikzpicture}
                \caption{The periodic quiver for $\mathbb{C}^{2}$ and its perfect matchings, represented here as orientations of the quiver.}
                \label{c2_quiver_pm}
            \end{figure}
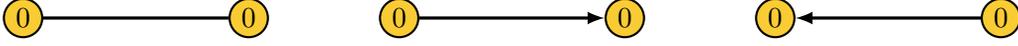

Algebraic dimensional reduction is indeed a specific instance of products and corresponds to the product of the quiver theory for $\mathrm{CY}_{m+2}$ with the simplest $m=0$ quiver theory, the one for $\mathbb{C}^{2}$. This theory is shown in \fref{c2_quiver_pm} and has two perfect matchings. We can use any of them and get the same result. Similarly any perfect matching used for the $\mathrm{CY}_{m+2}$ theory gives the same quiver theory for $\mathrm{CY}_{m+2}\times \mathbb{C}$ up to a relabeling of fields.
 
 \subsection{Orbifold Reduction}
    
\begin{figure}[ht]
            \centering
            \includegraphics[width=\textwidth]{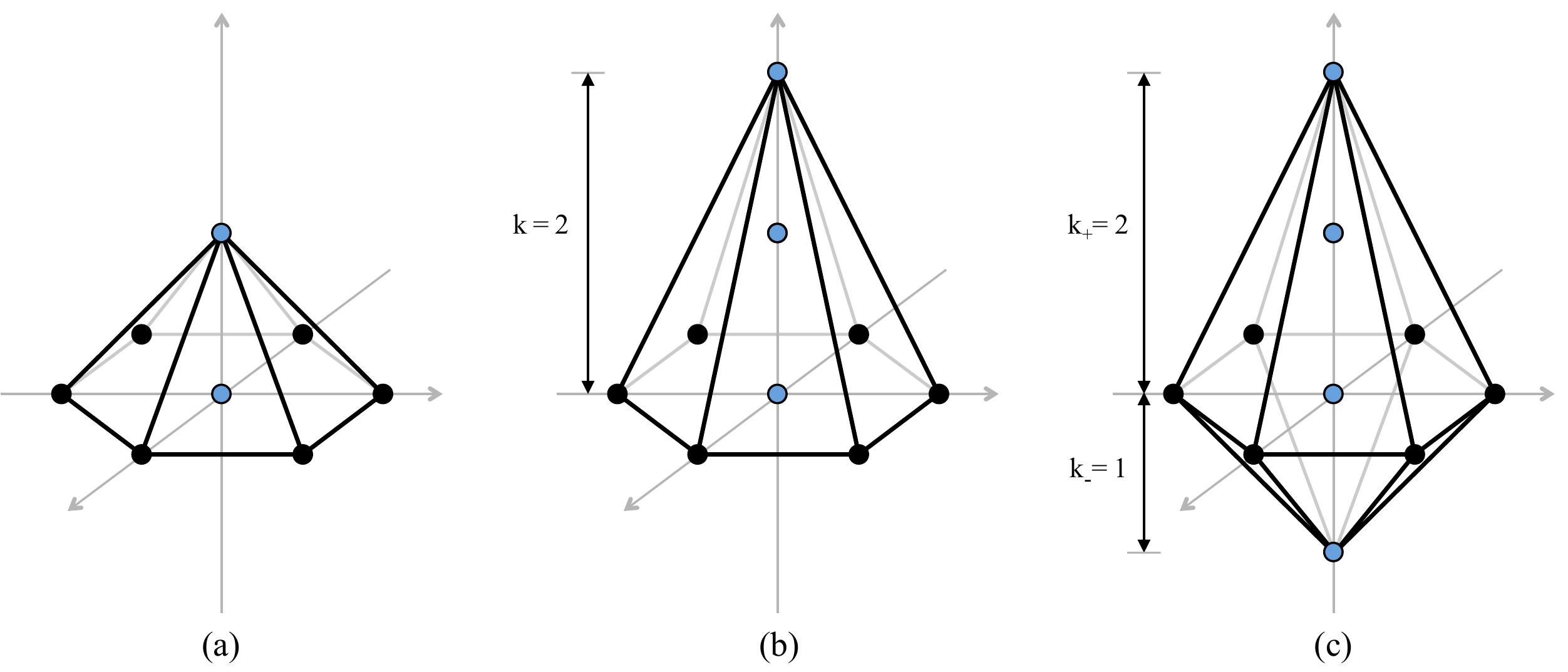}
        \caption{Toric diagrams for: a) the dimensional reduction of $dP_3$ to $dP_3\times \mathbb{C}$, b) a $(dP_3\times \mathbb{C})/\mathbb{Z}_k$ orbifold with $k=2$ and c) an orbifold reduction of $dP_3$ with $k_+=2$ and $k_-=1$.}
            \label{red_orb_orbred}
        \end{figure}

Orbifold reduction is a generalization of dimensional reduction that constructs a quiver theory for a toric $\mathrm{CY}_{4}$ from a that of a toric $\mathrm{CY}_{3}$ \cite{Franco:2016fxm}.\footnote{This corresponds to going from $m=1$ to $m=2$. The procedure can be naturally extended to higher $m$.} It adds a third dimension to the toric diagram $T_{\mathrm{CY}_{3}}$ by adding images of one of its points up to some height $k_{+}$ above the central plane containing the $T_{\mathrm{CY}_{3}}$ and some depth $k_{-}$ below it (see \fref{red_orb_orbred}).

This process again corresponds to a specific case of a product. The orbifold reduction of a $4d$ $\mathcal{N} = 1$ quiver theory with periodic quiver $P$ using a perfect matching $q$ corresponds to the product $P_{p} \times A^{(k_{+}+k_{-})}_{q}$. Here $A^{(k)}$ is the $6d$ $\mathcal{N} = (1,0)$ quiver theory for $\mathbb{C}^{2}/\mathbb{Z}_{k}$, i.e. the affine necklace quiver of type $A$ with $k$ nodes. A perfect matching of an $m=0$ quiver is just a choice of orientation of its edges, so the perfect matching $p$ is such that $k_{+}$ arrows point up while $k_{-}$ arrows point down. There are $\binom{k_{+} + k_{-}}{k_{-}}$ such perfect matchings. They all realize theories corresponding to the same geometry and are related by a sequence of trialities.

\subsection{3d Printing}  

        \label{subsection:product_vs_printing}

Another algorithm for efficiently constructing quiver theories for toric CYs starting from simpler parent geometries is {\it $3d$ printing}. $3d$ printing allows one to add images of multiple points in the toric diagram (we refer to \cite{Franco:2018qsc} for details). $3d$ printing is indeed more general than the CY product in two senses:
        \begin{itemize}
             \item
                 All the geometries that can be addressed with CY products can also be reached by a sequence of 3d printings that increase $m$ by one at a time. The converse is not true; there are geometries that can be realized by $3d$ printing but not as CY products. The simplest such example is the conifold. As it is evident from its toric diagram, shown in \fref{toric_diagram_quiver_conifold}, it can be constructed by lifting both the points in the toric diagram of $\mathbb{C}^{2}$. On the other hand, it is clear that it is not possible to produce it by a product.
            \item
                Even if the same geometry can be realized by both processes, there might be phases of the quiver theories that can be obtained via $3d$ printing but not via a product. A simple example of this phenomenon is $F_{0}$. Phase 2 of $F_{0}$ can be obtained using either construction but only $3d$ printing is able to construct phase 1. 
         \end{itemize}

           Despite these relative disadvantages, the CY product is a superior method for geometries that can be reached via both methods for several reasons:
         \begin{itemize}
             \item
The CY product is much more efficient. This is true even for simple geometries. As an example, let us consider the construction of a quiver theory for the conifold$\times$conifold. In order to $3d$ print this theory starting from the conifold, we first need to produce an intermediate CY 4-fold that is the dimensional reduction of the conifold, i.e. conifold$\times \mathbb{C}$. Then two points of its toric diagram must be lifted to produce the conifold$\times$conifold. To carry out this process we will have to compute the perfect matchings not only for the conifold but also for the intermediate conifold$\times \mathbb{C}$ theory. The difficulties of constructing the necessary quiver blocks and computing perfect matchings at every intermediate step makes $3d$ printing impractical if the difference between the dimensions of the input and target geometries is large.
            \item
                The CY product always produces {\it reduced} theories, which is not the case with $3d$ printing which often results in reducible, also known as inconsistent, theories which need to be reduced \cite{Franco:2018qsc}. 
            \item
                Unlike $3d$ printing the CY product does not generate mass terms in the superpotential. This not only reduces the computational burden but it also means that CY product provides a more direct way of arriving to the final quiver theory, without the need to integrate out massive fields at the end.
            \item
                More importantly, in addition to these computational advantages, the CY product provides us with a concise and much clearer relationship between the input and target geometries. This becomes more striking as the difference between the dimensions of the input and target geometries increases.
         \end{itemize}

Having considered the relative merits of the two constructions we turn to some speculation about their relation. While we have restricted ourselves to the case in which the periodic quivers for both theories are embedded in tori, more generally we can regard the product construction as a method for producing a quiver embedded in $S\times T$ given two quivers embedded in manifolds $S$ and $T$. We can also consider cases where the manifolds have a boundary. Imagine $T$ has a boundary $\partial T$. In that case the resulting quiver will be embedded in a manifold $S\times T$ with boundary $S\times \partial T$. Arguably the simplest case of this situation is when $T$ is the line segment $I$. The basic building block of $3d$ printing, a quiver block $\mathcal{Q}^{(m+1)}_{p}$, is a graph embedded in $\mathbb{T}^{m}\times I$ and indeed can be regarded as a product of an $m$-graded periodic quiver $Q^{(m)}$ using a perfect matching $p$ with a simple quiver embedded in a line segment as follows\footnote{The notation in the figure is inspired by the one used for quiver blocks in $3d$ printing in \cite{Franco:2018qsc}. In that context, the nodes $\overline{\star}$ and $\underline{\star}$ would correspond to the two images of a node $\star$ at the two endpoints of a line segment.} 
    \begin{align}
        \mathcal{Q}^{(m+1)}_{p} =  Q^{(m)}_{p} \times \mathord{\begin{tikzpicture}[baseline = 4ex,scale = 1.75]
                    \tikzstyle{every node}=[circle,thick,fill=yellow2,draw,inner sep=1.25pt,font=\scriptsize]
                    \draw (0,0) node(a){$\underline{\star}$};
                    \draw (0,1) node(b){$\overline{\star}$};
                    \draw [very thick,-latex] (a) -- (b);
                \end{tikzpicture}}
    \end{align}
As usual, we have indicated the perfect matching of the $m=0$ quiver by specifying an orientation of its fields. This construction realizes both the field content and the superpotential of the quiver block. 

It is therefore natural to expect that $3d$ printing and product are two instances of a single overarching construction. Such procedure would include both the products of $m$-graded quivers embedded in manifolds, possibly with boundaries, and an operation to glue two such manifolds along their boundaries under suitable conditions. We leave the task of understanding this construction in complete generality and its physical realization to a future work.

\section{Conclusions}

\label{section_conclusions}

Over the years, there has been tremendous progress in the map between the geometry of singularities and the corresponding quiver theories on branes. This started with a few isolated examples of CY 3-folds and evolved into the development of brane tilings, tools that vastly simplify that study of infinite classes of geometries. Similar tools were later developed for higher dimensional CYs. We regard the CY product as a significant development in the arsenal of tools to connect geometry and quiver theories. It allows us to straightforwardly compute quiver theories in cases that were previously out of practical reach.

We envision multiple directions for future research. To name a few:

\begin{itemize}

\item The CY product will help investigating the order $(m+1)$ dualities of the $m$-graded quiver theories associated to CY $(m+2)$-folds. There is a large amount of freedom in this construction: choice of phases for the quiver theories of the parent geometries and choice of perfect matchings for the interlacing points.\footnote{Moreover, the perfect matchings are phase dependent.} Therefore, given a target CY, there are multiple possible decompositions into CY factors. In fact, different decompositions can even differ in the dimension of the components. It is therefore worthwhile to study the interplay between this vast landscape of possibilities and the intricate space of dual theories.

\item The CY product is particularly amenable to automatic computer implementation. It is therefore ideally suited for generating large datasets of CYs/quiver theories. Such datasets would provide valuable insights into the structure of these theories. Moreover, they can be used to test the applicability of modern ideas such as machine learning to problems involving quiver theories, such as the classification of duals for general $m$. Initial explorations of these ideas have been undertaken in \cite{toappear0}. 

\item As mentioned in \sref{subsec_F0} in the case of $F_0^{(m)}$, the CY product can be applied iteratively, equivalently using multiple factors. In this way, it is possible to build quiver theories for complicated, higher dimensional geometries using very simple, low dimensional building blocks. A similar approach has been exploited to build some of the infinite classes of theories in \cite{Closset:2018axq}.

\item From a first principle perspective, we can calculate the quivers associated with a CY$_{m+2}$ via the topological B-model \cite{Aspinwall:2008jk,lam2014calabi,Franco:2017lpa,Closset:2018axq}. However, this approach requires knowledge of the fractional branes as a starting point, which is often challenging.  It would be interesting to investigate the correspondence between the B-model and CY product approaches.

\end{itemize}

\acknowledgments
 
We would like to thank C. Closset and G. Musiker for enjoyable discussions and related collaborations. The research of SF was supported by the U.S. National Science Foundation grants PHY-1820721 and DMS-1854179. AH was supported by INFN grant GSS (Gauge Theories, Strings and Supergravity).

\appendix

\section{Some Details About $\mathcal{W}_{PQ}$}
    \label{mixed_potential_terms_appendix}

In this appendix we expand our discussion of the mixed terms $\mathcal{W}_{PQ}$ that we introduced in \sref{section_superpotential}. For simplicity, let us first consider the next to simplest case, namely terms coming from a quartic term $T_P$ and a cubic term $T_Q$:
\beq
        T_{P} = X_{i_{1}i_{2}}^{(c_{1})}X_{i_{2}i_{3}}^{(c_{2})}X_{i_{3}i_{4}}^{(c_{3})}\bar{X}^{(m-1-c_{1}-c_{2}-c_{3})}_{i_{4}i_{1}} \ \ \ , \ \ \ T_{Q} = Y_{j_{1}j_{2}}^{(d_{1})}Y_{j_{2}j_{3}}^{(d_{2})}\bar{Y}_{j_{3}j_{1}}^{(n-1-d_{1}-d_{2})} ~.  
\eeq

We can reduce the order of the terms in $T_P$ by introducing two auxiliary massive fields $M_{i_{1}i_{3}}^{(c_{1}+c_{2})}$ and $\bar{M}_{i_{3}i_{1}}^{(m-1-c_{2}-c_{2})}$, with the following superpotential
\beq
        C_{P} = X_{i_{1}i_{2}}^{(c_{1})}X_{i_{2}i_{3}}^{(c_{2})}\bar{M}_{i_{3}i_{1}}^{(m-1-c_{1}-c_{2})} + M_{i_{1}i_{3}}^{(c_{1}+c_{2})}X_{i_{3}i_{4}}^{(c_{3})}\bar{X}^{(m-1-c_{1}-c_{2}-c_{3})}_{i_{4}i_{1}} - M_{i_{1}i_{3}}^{(c_{1}+c_{2})}\bar{M}_{i_{3}i_{1}}^{(m - c_{1}-c_{2})} ~.
\eeq
It is straightforward to verify that integrating out the two massive fields, $C_{P}$ gives back the quartic term $T_{P}$. 

\begin{figure}[ht]
        \centering
        \begin{subfigure}[c]{0.32\textwidth} 
        \centering
        \begin{tikzpicture}[scale=0.5 , decoration={markings,mark=at position \arrowHeadPosition with {\arrow{latex}}}] 
            \tikzstyle{every node}=[circle,thick,fill=yellow2,draw,inner sep=1.5pt,font=\tiny]
                \draw[step=1.0,gray,very thin] (1,1) grid (5,4);
                \draw (1,1) node(n1)[draw = none , fill = none,left]{$j_{1}$};
                \draw (1,2) node(n2)[draw = none , fill = none,left]{$j_{2}$};
                \draw (1,3) node(n3)[draw = none , fill = none,left]{$j_{3}$};
                \draw (1,4) node(n4)[draw = none , fill = none,left]{$j_{1}$};
                \draw (1,1) node(n1)[draw = none , fill = none,below]{$i_{1}$};
                \draw (2,1) node(n2)[draw = none , fill = none,below]{$i_{2}$};
                \draw (3,1) node(n3)[draw = none , fill = none,below]{$i_{3}$};
                \draw (4,1) node(n4)[draw = none , fill = none,below]{$i_{4}$};
                \draw (5,1) node(n5)[draw = none , fill = none,below]{$i_{1}$};
                \draw (4,3) node(n1){};
                \draw (1,1) node(n2){};
                \draw (2,1) node(n3){};
                \draw (3,2) node(n4){};
                \draw (5,4) node(n5){};
                \draw [postaction={decorate},red](n1) -- (n5);
                \draw [postaction={decorate} ](n2) -- (n3);
                \draw [postaction={decorate} ](n3) -- (n4);
                \draw [postaction={decorate} ](n4) -- (n1);
        \end{tikzpicture} 
        \end{subfigure}
        \begin{subfigure}[c]{0.32\textwidth} 
        \centering
        \begin{tikzpicture}[scale=0.5 , decoration={markings,mark=at position \arrowHeadPosition with {\arrow{latex}}}] 
            \tikzstyle{every node}=[circle,thick,fill=yellow2,draw,inner sep=1.5pt,font=\tiny]
                \draw[step=1.0,gray,very thin] (1,1) grid (5,4);
                \draw (1,1) node(n1)[draw = none , fill = none,left]{$j_{1}$};
                \draw (1,2) node(n2)[draw = none , fill = none,left]{$j_{2}$};
                \draw (1,3) node(n3)[draw = none , fill = none,left]{$j_{3}$};
                \draw (1,4) node(n4)[draw = none , fill = none,left]{$j_{1}$};
                \draw (1,1) node(n1)[draw = none , fill = none,below]{$i_{1}$};
                \draw (2,1) node(n2)[draw = none , fill = none,below]{$i_{2}$};
                \draw (3,1) node(n3)[draw = none , fill = none,below]{$i_{3}$};
                \draw (4,1) node(n4)[draw = none , fill = none,below]{$i_{4}$};
                \draw (5,1) node(n5)[draw = none , fill = none,below]{$i_{1}$};
                \draw (4,3) node(n1){};
                \draw (1,1) node(n2){};
                \draw (2,2) node(n3){};
                \draw (3,2) node(n4){};
                \draw (5,4) node(n5){};
                \draw [postaction={decorate},red](n1) -- (n5);
                \draw [postaction={decorate} ](n2) -- (n3);
                \draw [postaction={decorate} ](n3) -- (n4);
                \draw [postaction={decorate} ](n4) -- (n1);
        \end{tikzpicture} 
        \end{subfigure}
        \begin{subfigure}[c]{0.32\textwidth}
        \centering 
        \begin{tikzpicture}[scale=0.5 , decoration={markings,mark=at position \arrowHeadPosition with {\arrow{latex}}}] 
            \tikzstyle{every node}=[circle,thick,fill=yellow2,draw,inner sep=1.5pt,font=\tiny]
                \draw[step=1.0,gray,very thin] (1,1) grid (5,4);
                \draw (1,1) node(n1)[draw = none , fill = none,left]{$j_{1}$};
                \draw (1,2) node(n2)[draw = none , fill = none,left]{$j_{2}$};
                \draw (1,3) node(n3)[draw = none , fill = none,left]{$j_{3}$};
                \draw (1,4) node(n4)[draw = none , fill = none,left]{$j_{1}$};
                \draw (1,1) node(n1)[draw = none , fill = none,below]{$i_{1}$};
                \draw (2,1) node(n2)[draw = none , fill = none,below]{$i_{2}$};
                \draw (3,1) node(n3)[draw = none , fill = none,below]{$i_{3}$};
                \draw (4,1) node(n4)[draw = none , fill = none,below]{$i_{4}$};
                \draw (5,1) node(n5)[draw = none , fill = none,below]{$i_{1}$};
                \draw (3,3) node(n1){};
                \draw (4,3) node(n2){};
                \draw (1,1) node(n3){};
                \draw (2,2) node(n4){};
                \draw (5,4) node(n5){};
                \draw [postaction={decorate} ](n1) -- (n2);
                \draw [postaction={decorate},red ](n2) -- (n5);
                \draw [postaction={decorate} ](n3) -- (n4);
                \draw [postaction={decorate} ](n4) -- (n1);
        \end{tikzpicture} 
        \end{subfigure}
        \caption{The three terms in $\mathcal{W}_{PQ}$ coming from a quartic $T_{P}$ and a cubic $T_{Q}$. Red arrows represent the products of a field in $p$ and a field in $q$. Black arrows descend from fields that are not in $p$ or $q$.}
        \label{cubic_quartic_potential}
    \end{figure}
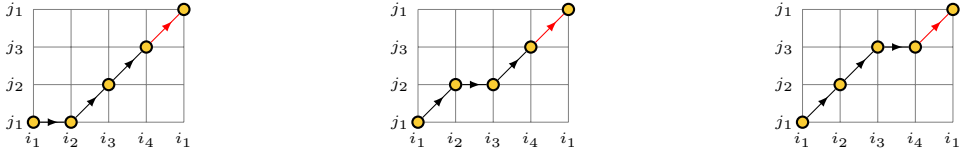

It is now easy to construct the terms in the product superpotential coming from $C_{P}$ and $T_{Q}$. After integrating out the massive fields, most of the terms correspond to those in $\mathcal{W}_P$ due to $T_{P}$, $\mathcal{W}_Q$ due to $T_{Q}$ or in $\mathcal{W}_C$ due to fields in $T_{P}$ and $T_{Q}$. In addition, we get three extra quartic terms coming from the contributions of $T_{P}$ and $T_{Q}$ to $\mathcal{W}_{PQ}$.  
    These terms are:
    \begin{align}
         &(-1)^{m+n+1+c_{2}+d_{1}}Z_{(i_{1},j_{1})(i_{2},j_{1})}^{(c_{1})}Z_{(i_{2},j_{1}),(i_{3},j_{2})}^{(c_{2}+d_{1}+1)}Z^{(c_{3}+d_{2}+1)}_{(i_{3},j_{2}),(i_{4},j_{3})}\bar{Z}^{(n+m-2-c_{2}-c_{3}-d_{1}-d_{2})}_{(i_{4},j_{3}),(i_{1},j_{1})} \nonumber\\
         +&(-1)^{m+n+1+c_{1}+c_{2}+d_{1}}Z_{(i_{1},j_{1})(i_{2},j_{2})}^{(c_{1}+d_{1}+1)}Z_{(i_{2},j_{2}),(i_{3},j_{2})}^{(c_{2})}Z^{(c_{3}+d_{2}+1)}_{(i_{3},j_{2}),(i_{4},j_{3})}\bar{Z}^{(n+m-2-c_{1}-c_{3}-d_{1}-d_{2})}_{(i_{4},j_{3}),(i_{1},j_{1})} \nonumber \\
         +&(-1)^{m+n+c_{1}+d_{1}}Z_{(i_{1},j_{1})(i_{2},j_{2})}^{(c_{1}+d_{1}+1)}Z_{(i_{2},j_{2}),(i_{3},j_{3})}^{(c_{2}+d_{2}+1)}Z^{(c_{3})}_{(i_{3},j_{3}),(i_{4},j_{3})}\bar{Z}^{(n+m-2-c_{2}-c_{3}-d_{1}-d_{2})}_{(i_{4},j_{3}),(i_{1},j_{1})} 
\label{3_terms_WPQ}  
  \end{align}
    These terms are depicted graphically in \fref{cubic_quartic_potential}, which shows them on a torus whose fundamental cycles are the two terms $T_{P}$ and $T_{Q}$.\footnote{Notice that these should not be confused with the fundamental cycles of the periodic quivers.}

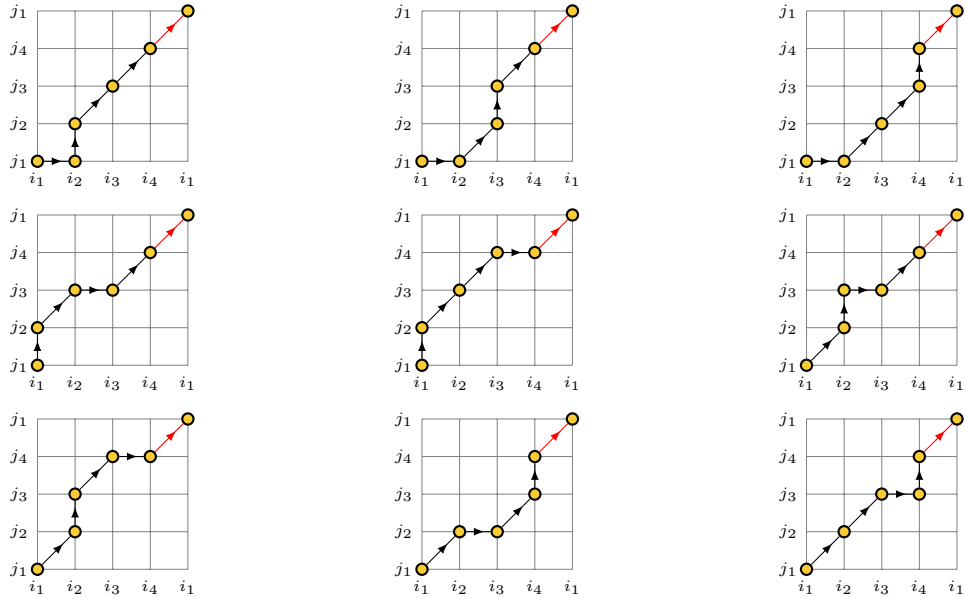
\begin{figure}
        \centering
        \begin{subfigure}[c]{0.32\textwidth} 
        \centering
            \begin{tikzpicture}[scale=0.5 , decoration={markings,mark=at position \arrowHeadPosition with {\arrow{latex}}}] 
                \tikzstyle{every node}=[circle,thick,fill=yellow2,draw,inner sep=1.5pt,font=\tiny]
                    \draw[step=1.0,gray,very thin] (1,1) grid (5,5);
                    \draw (1,1) node(n1)[draw = none , fill = none,left]{$j_{1}$};
                    \draw (1,2) node(n2)[draw = none , fill = none,left]{$j_{2}$};
                    \draw (1,3) node(n3)[draw = none , fill = none,left]{$j_{3}$};
                    \draw (1,4) node(n4)[draw = none , fill = none,left]{$j_{4}$};
                    \draw (1,5) node(n4)[draw = none , fill = none,left]{$j_{1}$};
                    \draw (1,1) node(n1)[draw = none , fill = none,below]{$i_{1}$};
                    \draw (2,1) node(n2)[draw = none , fill = none,below]{$i_{2}$};
                    \draw (3,1) node(n3)[draw = none , fill = none,below]{$i_{3}$};
                    \draw (4,1) node(n4)[draw = none , fill = none,below]{$i_{4}$};
                    \draw (5,1) node(n5)[draw = none , fill = none,below]{$i_{1}$};
                    \draw (4,4) node(n1){};
                    \draw (1,1) node(n2){};
                    \draw (2,1) node(n3){};
                    \draw (2,2) node(n4){};
                    \draw (3,3) node(n5){};
                    \draw (5,5) node(n6){};
                    \draw [postaction={decorate},red](n1) -- (n6);
                    \draw [postaction={decorate} ](n2) -- (n3);
                    \draw [postaction={decorate} ](n3) -- (n4);
                    \draw [postaction={decorate} ](n4) -- (n5);
                    \draw [postaction={decorate} ](n5) -- (n1);
            \end{tikzpicture} 
        \end{subfigure}
        \begin{subfigure}[c]{0.32\textwidth} 
        \centering
            \begin{tikzpicture}[scale=0.5 , decoration={markings,mark=at position \arrowHeadPosition with {\arrow{latex}}}] 
                \tikzstyle{every node}=[circle,thick,fill=yellow2,draw,inner sep=1.5pt,font=\tiny]
                    \draw[step=1.0,gray,very thin] (1,1) grid (5,5);
                    \draw (1,1) node(n1)[draw = none , fill = none,left]{$j_{1}$};
                    \draw (1,2) node(n2)[draw = none , fill = none,left]{$j_{2}$};
                    \draw (1,3) node(n3)[draw = none , fill = none,left]{$j_{3}$};
                    \draw (1,4) node(n4)[draw = none , fill = none,left]{$j_{4}$};
                    \draw (1,5) node(n4)[draw = none , fill = none,left]{$j_{1}$};
                    \draw (1,1) node(n1)[draw = none , fill = none,below]{$i_{1}$};
                    \draw (2,1) node(n2)[draw = none , fill = none,below]{$i_{2}$};
                    \draw (3,1) node(n3)[draw = none , fill = none,below]{$i_{3}$};
                    \draw (4,1) node(n4)[draw = none , fill = none,below]{$i_{4}$};
                    \draw (5,1) node(n5)[draw = none , fill = none,below]{$i_{1}$};
                    \draw (4,4) node(n1){};
                    \draw (1,1) node(n2){};
                    \draw (2,1) node(n3){};
                    \draw (3,2) node(n4){};
                    \draw (3,3) node(n5){};
                    \draw (5,5) node(n6){};
                    \draw [postaction={decorate},red](n1) -- (n6);
                    \draw [postaction={decorate} ](n2) -- (n3);
                    \draw [postaction={decorate} ](n3) -- (n4);
                    \draw [postaction={decorate} ](n4) -- (n5);
                    \draw [postaction={decorate} ](n5) -- (n1);
            \end{tikzpicture} 
        \end{subfigure}
        \begin{subfigure}[c]{0.32\textwidth} 
        \centering
            \begin{tikzpicture}[scale=0.5 , decoration={markings,mark=at position \arrowHeadPosition with {\arrow{latex}}}] 
                \tikzstyle{every node}=[circle,thick,fill=yellow2,draw,inner sep=1.5pt,font=\tiny]
                    \draw[step=1.0,gray,very thin] (1,1) grid (5,5);
                    \draw (1,1) node(n1)[draw = none , fill = none,left]{$j_{1}$};
                    \draw (1,2) node(n2)[draw = none , fill = none,left]{$j_{2}$};
                    \draw (1,3) node(n3)[draw = none , fill = none,left]{$j_{3}$};
                    \draw (1,4) node(n4)[draw = none , fill = none,left]{$j_{4}$};
                    \draw (1,5) node(n4)[draw = none , fill = none,left]{$j_{1}$};
                    \draw (1,1) node(n1)[draw = none , fill = none,below]{$i_{1}$};
                    \draw (2,1) node(n2)[draw = none , fill = none,below]{$i_{2}$};
                    \draw (3,1) node(n3)[draw = none , fill = none,below]{$i_{3}$};
                    \draw (4,1) node(n4)[draw = none , fill = none,below]{$i_{4}$};
                    \draw (5,1) node(n5)[draw = none , fill = none,below]{$i_{1}$};
                    \draw (4,3) node(n1){};
                    \draw (4,4) node(n2){};
                    \draw (1,1) node(n3){};
                    \draw (2,1) node(n4){};
                    \draw (3,2) node(n5){};
                    \draw (5,5) node(n6){};
                    \draw [postaction={decorate}](n1) -- (n2);
                    \draw [postaction={decorate},red](n2) -- (n6);
                    \draw [postaction={decorate} ](n3) -- (n4);
                    \draw [postaction={decorate} ](n4) -- (n5);
                    \draw [postaction={decorate} ](n5) -- (n1);
            \end{tikzpicture} 
        \end{subfigure}

        \begin{subfigure}[c]{0.32\textwidth} 
        \centering
            \begin{tikzpicture}[scale=0.5 , decoration={markings,mark=at position \arrowHeadPosition with {\arrow{latex}}}] 
                \tikzstyle{every node}=[circle,thick,fill=yellow2,draw,inner sep=1.5pt,font=\tiny]
                    \draw[step=1.0,gray,very thin] (1,1) grid (5,5);
                    \draw (1,1) node(n1)[draw = none , fill = none,left]{$j_{1}$};
                    \draw (1,2) node(n2)[draw = none , fill = none,left]{$j_{2}$};
                    \draw (1,3) node(n3)[draw = none , fill = none,left]{$j_{3}$};
                    \draw (1,4) node(n4)[draw = none , fill = none,left]{$j_{4}$};
                    \draw (1,5) node(n4)[draw = none , fill = none,left]{$j_{1}$};
                    \draw (1,1) node(n1)[draw = none , fill = none,below]{$i_{1}$};
                    \draw (2,1) node(n2)[draw = none , fill = none,below]{$i_{2}$};
                    \draw (3,1) node(n3)[draw = none , fill = none,below]{$i_{3}$};
                    \draw (4,1) node(n4)[draw = none , fill = none,below]{$i_{4}$};
                    \draw (5,1) node(n5)[draw = none , fill = none,below]{$i_{1}$};
                    \draw (4,4) node(n1){};
                    \draw (1,1) node(n2){};
                    \draw (1,2) node(n3){};
                    \draw (2,3) node(n4){};
                    \draw (3,3) node(n5){};
                    \draw (5,5) node(n6){};
                    \draw [postaction={decorate},red](n1) -- (n6);
                    \draw [postaction={decorate} ](n2) -- (n3);
                    \draw [postaction={decorate} ](n3) -- (n4);
                    \draw [postaction={decorate} ](n4) -- (n5);
                    \draw [postaction={decorate} ](n5) -- (n1);
            \end{tikzpicture} 
        \end{subfigure}
        \begin{subfigure}[c]{0.32\textwidth} 
        \centering
            \begin{tikzpicture}[scale=0.5 , decoration={markings,mark=at position \arrowHeadPosition with {\arrow{latex}}}] 
                    \tikzstyle{every node}=[circle,thick,fill=yellow2,draw,inner sep=1.5pt,font=\tiny]
                        \draw[step=1.0,gray,very thin] (1,1) grid (5,5);
                        \draw (1,1) node(n1)[draw = none , fill = none,left]{$j_{1}$};
                        \draw (1,2) node(n2)[draw = none , fill = none,left]{$j_{2}$};
                        \draw (1,3) node(n3)[draw = none , fill = none,left]{$j_{3}$};
                        \draw (1,4) node(n4)[draw = none , fill = none,left]{$j_{4}$};
                        \draw (1,5) node(n4)[draw = none , fill = none,left]{$j_{1}$};
                        \draw (1,1) node(n1)[draw = none , fill = none,below]{$i_{1}$};
                        \draw (2,1) node(n2)[draw = none , fill = none,below]{$i_{2}$};
                        \draw (3,1) node(n3)[draw = none , fill = none,below]{$i_{3}$};
                        \draw (4,1) node(n4)[draw = none , fill = none,below]{$i_{4}$};
                        \draw (5,1) node(n5)[draw = none , fill = none,below]{$i_{1}$};
                        \draw (3,4) node(n1){};
                        \draw (4,4) node(n2){};
                        \draw (1,1) node(n3){};
                        \draw (1,2) node(n4){};
                        \draw (2,3) node(n5){};
                        \draw (5,5) node(n6){};
                        \draw [postaction={decorate}](n1) -- (n2);
                        \draw [postaction={decorate},red](n2) -- (n6);
                        \draw [postaction={decorate} ](n3) -- (n4);
                        \draw [postaction={decorate} ](n4) -- (n5);
                        \draw [postaction={decorate} ](n5) -- (n1);
                \end{tikzpicture} 
            \end{subfigure}
        \begin{subfigure}[c]{0.32\textwidth} 
        \centering
            \begin{tikzpicture}[scale=0.5 , decoration={markings,mark=at position \arrowHeadPosition with {\arrow{latex}}}] 
                \tikzstyle{every node}=[circle,thick,fill=yellow2,draw,inner sep=1.5pt,font=\tiny]
                    \draw[step=1.0,gray,very thin] (1,1) grid (5,5);
                    \draw (1,1) node(n1)[draw = none , fill = none,left]{$j_{1}$};
                    \draw (1,2) node(n2)[draw = none , fill = none,left]{$j_{2}$};
                    \draw (1,3) node(n3)[draw = none , fill = none,left]{$j_{3}$};
                    \draw (1,4) node(n4)[draw = none , fill = none,left]{$j_{4}$};
                    \draw (1,5) node(n4)[draw = none , fill = none,left]{$j_{1}$};
                    \draw (1,1) node(n1)[draw = none , fill = none,below]{$i_{1}$};
                    \draw (2,1) node(n2)[draw = none , fill = none,below]{$i_{2}$};
                    \draw (3,1) node(n3)[draw = none , fill = none,below]{$i_{3}$};
                    \draw (4,1) node(n4)[draw = none , fill = none,below]{$i_{4}$};
                    \draw (5,1) node(n5)[draw = none , fill = none,below]{$i_{1}$};
                    \draw (4,4) node(n1){};
                    \draw (1,1) node(n2){};
                    \draw (2,2) node(n3){};
                    \draw (2,3) node(n4){};
                    \draw (3,3) node(n5){};
                    \draw (5,5) node(n6){};
                    \draw [postaction={decorate},red](n1) -- (n6);
                    \draw [postaction={decorate} ](n2) -- (n3);
                    \draw [postaction={decorate} ](n3) -- (n4);
                    \draw [postaction={decorate} ](n4) -- (n5);
                    \draw [postaction={decorate} ](n5) -- (n1);
            \end{tikzpicture} 
        \end{subfigure}

        \begin{subfigure}[c]{0.32\textwidth} 
        \centering
            \begin{tikzpicture}[scale=0.5 , decoration={markings,mark=at position \arrowHeadPosition with {\arrow{latex}}}] 
                \tikzstyle{every node}=[circle,thick,fill=yellow2,draw,inner sep=1.5pt,font=\tiny]
                    \draw[step=1.0,gray,very thin] (1,1) grid (5,5);
                    \draw (1,1) node(n1)[draw = none , fill = none,left]{$j_{1}$};
                    \draw (1,2) node(n2)[draw = none , fill = none,left]{$j_{2}$};
                    \draw (1,3) node(n3)[draw = none , fill = none,left]{$j_{3}$};
                    \draw (1,4) node(n4)[draw = none , fill = none,left]{$j_{4}$};
                    \draw (1,5) node(n4)[draw = none , fill = none,left]{$j_{1}$};
                    \draw (1,1) node(n1)[draw = none , fill = none,below]{$i_{1}$};
                    \draw (2,1) node(n2)[draw = none , fill = none,below]{$i_{2}$};
                    \draw (3,1) node(n3)[draw = none , fill = none,below]{$i_{3}$};
                    \draw (4,1) node(n4)[draw = none , fill = none,below]{$i_{4}$};
                    \draw (5,1) node(n5)[draw = none , fill = none,below]{$i_{1}$};
                    \draw (3,4) node(n1){};
                    \draw (4,4) node(n2){};
                    \draw (1,1) node(n3){};
                    \draw (2,2) node(n4){};
                    \draw (2,3) node(n5){};
                    \draw (5,5) node(n6){};
                    \draw [postaction={decorate} ](n1) -- (n2);
                    \draw [postaction={decorate},red](n2) -- (n6);
                    \draw [postaction={decorate} ](n3) -- (n4);
                    \draw [postaction={decorate} ](n4) -- (n5);
                    \draw [postaction={decorate} ](n5) -- (n1);
            \end{tikzpicture} 
        \end{subfigure}
            \begin{subfigure}[c]{0.32\textwidth} 
            \centering
                \begin{tikzpicture}[scale=0.5 , decoration={markings,mark=at position \arrowHeadPosition with {\arrow{latex}}}] 
                    \tikzstyle{every node}=[circle,thick,fill=yellow2,draw,inner sep=1.5pt,font=\tiny]
                        \draw[step=1.0,gray,very thin] (1,1) grid (5,5);
                        \draw (1,1) node(n1)[draw = none , fill = none,left]{$j_{1}$};
                        \draw (1,2) node(n2)[draw = none , fill = none,left]{$j_{2}$};
                        \draw (1,3) node(n3)[draw = none , fill = none,left]{$j_{3}$};
                        \draw (1,4) node(n4)[draw = none , fill = none,left]{$j_{4}$};
                        \draw (1,5) node(n4)[draw = none , fill = none,left]{$j_{1}$};
                        \draw (1,1) node(n1)[draw = none , fill = none,below]{$i_{1}$};
                        \draw (2,1) node(n2)[draw = none , fill = none,below]{$i_{2}$};
                        \draw (3,1) node(n3)[draw = none , fill = none,below]{$i_{3}$};
                        \draw (4,1) node(n4)[draw = none , fill = none,below]{$i_{4}$};
                        \draw (5,1) node(n5)[draw = none , fill = none,below]{$i_{1}$};
                        \draw (4,3) node(n1){};
                        \draw (4,4) node(n2){};
                        \draw (1,1) node(n3){};
                        \draw (2,2) node(n4){};
                        \draw (3,2) node(n5){};
                        \draw (5,5) node(n6){};
                        \draw [postaction={decorate} ](n1) -- (n2);
                        \draw [postaction={decorate},red](n2) -- (n6);
                        \draw [postaction={decorate} ](n3) -- (n4);
                        \draw [postaction={decorate} ](n4) -- (n5);
                        \draw [postaction={decorate} ](n5) -- (n1);
                \end{tikzpicture} 
            \end{subfigure}
            \begin{subfigure}[c]{0.32\textwidth} 
            \centering
                \begin{tikzpicture}[scale=0.5 , decoration={markings,mark=at position \arrowHeadPosition with {\arrow{latex}}}] 
                    \tikzstyle{every node}=[circle,thick,fill=yellow2,draw,inner sep=1.5pt,font=\tiny]
                        \draw[step=1.0,gray,very thin] (1,1) grid (5,5);
                        \draw (1,1) node(n1)[draw = none , fill = none,left]{$j_{1}$};
                        \draw (1,2) node(n2)[draw = none , fill = none,left]{$j_{2}$};
                        \draw (1,3) node(n3)[draw = none , fill = none,left]{$j_{3}$};
                        \draw (1,4) node(n4)[draw = none , fill = none,left]{$j_{4}$};
                        \draw (1,5) node(n4)[draw = none , fill = none,left]{$j_{1}$};
                        \draw (1,1) node(n1)[draw = none , fill = none,below]{$i_{1}$};
                        \draw (2,1) node(n2)[draw = none , fill = none,below]{$i_{2}$};
                        \draw (3,1) node(n3)[draw = none , fill = none,below]{$i_{3}$};
                        \draw (4,1) node(n4)[draw = none , fill = none,below]{$i_{4}$};
                        \draw (5,1) node(n5)[draw = none , fill = none,below]{$i_{1}$};
                        \draw (3,3) node(n1){};
                        \draw (4,3) node(n2){};
                        \draw (4,4) node(n3){};
                        \draw (1,1) node(n4){};
                        \draw (2,2) node(n5){};
                        \draw (5,5) node(n6){};
                        \draw [postaction={decorate}](n1) -- (n2);
                        \draw [postaction={decorate}](n2) -- (n3);
                        \draw [postaction={decorate},red](n3) -- (n6);
                        \draw [postaction={decorate} ](n4) -- (n5);
                        \draw [postaction={decorate} ](n5) -- (n1);
                \end{tikzpicture} 
            \end{subfigure}
            \caption{The 9 terms in $\mathcal{W}_{PQ}$ coming from both $T_{P}$ and $T_{Q}$ quartic. Red arrows represent the products of a field in $p$ and a field in $q$. Black arrows descend from fields that are not in $p$ or $q$.}
            \label{quartic_quartic_potential}
        \end{figure}

Similarly, we can go one step further and consider the case in which $T_{P}$ and $T_{Q}$ are quartic. Proceeding as before, we can integrate in massive fields, turning $T_{P}$ into a sum of cubic terms and a mass term. Next, we use the previous result for a quartic and cubic terms.\footnote{This procedure accounts to reducing both $T_{P}$ and $T_{Q}$ to cubic and mass terms by integrating in massive fields.} After integrating out the massive fields we obtain standard terms in $\mathcal{W}_P$, $\mathcal{W}_Q$ and $\mathcal{W}_C$. In addition, we get nine terms in $\mathcal{W}_{PQ}$, all of order 5. These terms are shown in \fref{quartic_quartic_potential}, which reveals an unexpected feature of the resulting terms. Surprisingly, they are not symmetric under the exchange of $T_{P}$ and $T_{Q}$. This can be seen by exchanging horizontal and vertical arrows in these terms. The images of three of the terms under this operation are absent in \fref{quartic_quartic_potential}. This might be puzzling at first sight, since the procedure we described seems to treat $T_{P}$ and $T_{Q}$ symmetrically. It turns out that the symmetry is actually broken by the order in which we integrate out the massive fields.

It may seem possible to restore the symmetry between $T_{P}$ and $T_{Q}$, i.e. between the horizontal and vertical directions, by adding the missing terms. However, there is no way to do this while satisfying the Kontsevich bracket condition. Therefore, in this case we are left with two choices, which lead to different superpotentials.\footnote{It would to interesting to see if and how this choice is present in the B-model computation of the superpotential. We suspect this is related to the choice of explicit representatives of cohomology classes needed for computations of the products $m_{k}$ with $k>2$.} It is natural to expect that these two theories are related by duality.

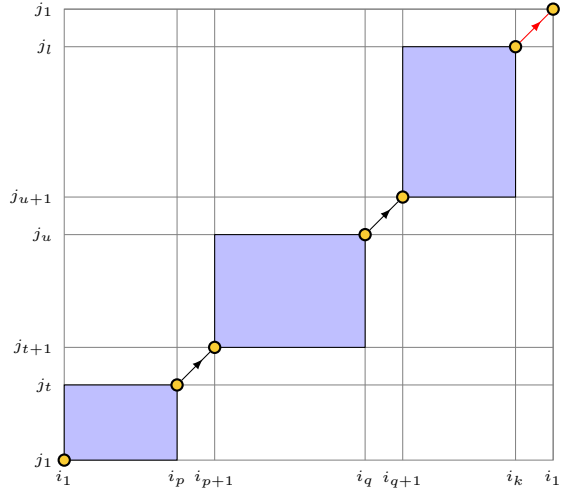
\begin{figure}
            \centering
            \begin{tikzpicture}[scale=0.5 , decoration={markings,mark=at position \arrowHeadPosition with {\arrow{latex}}}] 
                \tikzstyle{every node}=[font=\tiny]
                \draw (1,1) node(n1)[draw = none , fill = none,left]{$j_{1}$};
                \draw (1,3) node(n2)[draw = none , fill = none,left]{$j_{t}$};
                \draw (1,4) node(n3)[draw = none , fill = none,left]{$j_{t+1}$};
                \draw (1,7) node(n4)[draw = none , fill = none,left]{$j_{u}$};
                \draw (1,8) node(n4)[draw = none , fill = none,left]{$j_{u+1}$};
                \draw (1,12) node(n4)[draw = none , fill = none,left]{$j_{l}$};
                \draw (1,13) node(n4)[draw = none , fill = none,left]{$j_{1}$};
                \draw (1,1) node(n1)[draw = none , fill = none,below]{$i_{1}$};
                \draw (4,1) node(n2)[draw = none , fill = none,below]{$i_{p}$};
                \draw (5,1) node(n3)[draw = none , fill = none,below]{$i_{p+1}$};
                \draw (9,1) node(n4)[draw = none , fill = none,below]{$i_{q}$};
                \draw (10,1) node(n5)[draw = none , fill = none,below=0.25]{$i_{q+1}$};
                \draw (13,1) node(n2)[draw = none , fill = none,below]{$i_{k}$};
                \draw (14,1) node(n3)[draw = none , fill = none,below]{$i_{1}$};
                \tikzstyle{every node}=[circle,thick,fill=yellow2,draw,inner sep=1.5pt,font=\tiny]
                \draw[gray,very thin] (1,1) rectangle (14,13);
                \draw[gray,very thin] (4,1) -- (4,13);
                \draw[gray,very thin] (5,1) -- (5,13);
                \draw[gray,very thin] (9,1) -- (9,13);
                \draw[gray,very thin] (10,1) -- (10,13);
                \draw[gray,very thin] (13,1) -- (13,13);
                \draw[gray,very thin] (14,1) -- (14,13);
                \draw[gray,very thin] (1,3) -- (14,3);
                \draw[gray,very thin] (1,4) -- (14,4);
                \draw[gray,very thin] (1,7) -- (14,7);
                \draw[gray,very thin] (1,8) -- (14,8);
                \draw[gray,very thin] (1,12) -- (14,12);
                \draw[gray,very thin] (1,13) -- (14,13);
                \draw[fill = blue!25!white] (1,1) rectangle (4,3);
                \draw[fill = blue!25!white] (5,4) rectangle (9,7);
                \draw[fill = blue!25!white] (10,8) rectangle (13,12);
                \draw (1,1) node(n1){};
                \draw (4,3) node(n2){};
                \draw (5,4) node(n3){};
                \draw (9,7) node(n4){};
                \draw (10,8) node(n5){};
                \draw (13,12) node(n6){};
                \draw (14,13) node(n7){};
                \draw [postaction={decorate}](n2) -- (n3);
                \draw [postaction={decorate}](n4) -- (n5);
                \draw [postaction={decorate},red](n6) -- (n7);
            \end{tikzpicture}
            \caption{The general structure of a term in $\mathcal{W}_{PQ}$ descending from an order $k$ and an order $l$ terms. The blue boxes contain paths involving horizontal and vertical fields, i.e. products of a field and a node. The multiplicity of terms corresponds to the different ways of choosing the two black diagonal fields.}
            \label{general_kl_pontential}
        \end{figure}

Knowing the terms arising from an order $k-1$ term and an order $l$ term, we can recursively derive the terms arising from an order $k$ term and an order $l$ term. To do so, we can simply split the order $k$ term into an order $k-1$ term, a cubic term and a mass term. Continuing this iterative process for a few more steps we can infer the structure of the general case, which is depicted graphically in \fref{general_kl_pontential}. Every term in $\mathcal{W}_{PQ}$ contains one field that is the product of a field in $p$ and a field in $q$, and two fields that are the product of a field not in $p$ and a field not in $q$. They correspond the red and two black diagonal arrows. There is exactly one term for every choice of two diagonal black arrows. Every one of the three blue boxes contains a path between two of these fields composed exclusively of horizontal and vertical arrows, i.e. of fields that are the product of a field and a node. The precise path depends on the breaking of the order $k$ term into an order $k-1$ term, a cubic term and a mass term.  

\section{Products and Geometry}

\label{section_products_geometry}

Here we explain how the product theory gives rise to the desired geometry, which arises as its classical moduli space. To do so, we show how the perfect matchings of $P_{p} \times Q_{q}$ result in the toric diagram described by \eref{product_toric_diagram}.

First, we note that the collection of all the conjugated fields forms a perfect matching.\footnote{Recall that our convention is that the polarization of the $P_{p} \times Q_{q}$ quiver and hence the identity of the conjugated fields is determined by the choice of $p$ and $q$.} This is the perfect matching that corresponds to the ``central point" $(u_{0},v_{0})$ of $T_{\mathrm{CY}_{m+n+3}}$.

Given a perfect matching $\tilde{p}$ of $P$ we can construct a perfect matching that we will call $\tilde{p}\times q$ of $P_{p}\times Q_{q}$. If $\tilde{p}$ corresponds to the point $u_{i}$ in $T_{\mathrm{CY}_{m+2}}$ then $\tilde{p}\times q$ corresponds to the point $(u_{i},v_{0})$ of $T_{\mathrm{CY}_{m+n+3}}$. In order to construct $\tilde{p}\times q$, we divide the fields in $\tilde{p}$ into two sets. The first set $\tilde{p}_{0}$ contains the fields in $\tilde{p}$ that are also in $p$, while the second set $\tilde{p}_{*}$ contains the fields in $\tilde{p}$ that are not in $p$, namely
\beq
            \tilde{p}_{0} = \tilde{p} \cap p  \ \ \ \ \ , \ \ \ \ \ \tilde{p}_{*} = \tilde{p}\setminus p ~.
\eeq
        Then, $\tilde{p} \times q$ is
\beq
            \tilde{p} \times q = (I \times q)\cup (\tilde{p} \times J) \cup (\tilde{p}_{0} \times q) \cup (\tilde{p}_{*} \times \cancel{q})~,
\label{union_pm_product}
\eeq
where $\cancel{q}$ is the set of all fields in $Q$ that are not in $q$, i.e. it is the set of the conjugates of fields in $q$. Let us now define the sets that participate in the union \eref{union_pm_product}. The first two of these are defined as
        \begin{align}
            I \times q &= \left\{\bar{Z}^{(d+m+1)}_{(i,j_{1})(i,j_{2})}| i \in I , \bar{Y}^{(d)}_{j_{1},j_{2}} \in q \right\}  \nonumber\\
            \tilde{p} \times J &= \left\{\bar{Z}^{(c+n+1)}_{(i_{1},j)(i_{2},j)}|\bar{X}_{(i_{1}i_{2})}^{(c)} \in \tilde{p}_{0}, j \in J\right\} \cup \left\{Z^{(c)}_{(i_{1},j)(i_{2},j)}|X^{(c)}_{(i_{1},i_{2})} \in \tilde{p}_{*}, j\in J\right\}
        \end{align}
        i.e. $I \times q$ is just the set of fields that result from the product between a node $i$ of $P$ and a field in $q$, while $\tilde{p}\times J$ is the set of fields that result from the product between a field in $\tilde{p}$ and a node $j$ of $Q$. We have separated $\tilde{p} \times J$ into two pieces because the degree of the resulting field behaves differently depending on whether the original field is in $p$ or not.

        The set $p_{0} \times q$ is defined as
\beq
            \tilde{p}_{0} \times q = \left\{\bar{Z}^{(c+d)}_{(i_{1},j_{1})(i_{2},j_{2})}|\bar{X}_{(i_{1}i_{2})}^{(c)} \in \tilde{p}_{0}, \bar{Y}_{j_{1}j_{2}}^{(d)} \in q\right\} ~.
\eeq
This set has a simple interpretation: it consists of all the fields in $P_{p} \times Q_{p}$ that arise from a product between a field that is common to $p$ and $\tilde{p}$ and a field in $q$. 

The interpretation of $\tilde{p}_{*} \times \cancel{q}$ is similar. It consists of the fields that come from the product of a field that is in $\tilde{p}$ but not in $p$ with a field of $Q$ that is not in $q$, i.e.
   \beq
            \tilde{p}_{*} \times \cancel{q} = \left\{Z^{(c+d+1)}_{(i_{1},j_{1})(i_{2},j_{2})}|X^{(c)}_{i_{1}i_{2}} \in \tilde{p}_{*} , Y^{(d)}_{j_{1}j_{2}}\in \cancel{q}\right\} ~.
\eeq

Analogously, given a perfect matching $\tilde{q}$ of $Q$ corresponding to the point $v_{i}$ we can define a perfect matching $p\times \tilde{q}$ that corresponds to the point $(u_{0},v_{i})$ in $T_{\mathrm{CY}_{m+n+3}}$. It is defined as
\beq
            p \times \tilde{q} = (I \times \tilde{q})\cup (p \times J) \cup (p \times \tilde{q}_{0} )\cup (\cancel{p} \times \tilde{q}_{*}) ~.
\label{union_pm_product_2}
\eeq
As for $\tilde{p}$, we define $\tilde{q}_{0} = \tilde{q}\cap q$ and $\tilde{q}_{*} = \tilde{q}\setminus q$, while $\cancel{p}$ is the set of fields conjugate to those in $p$. The four sets in \eref{union_pm_product_2} are defined as follows 
        \begin{align}
             I \times \tilde{q} &= \left\{\bar{Z}^{(d+m+1)}_{(i,j_{1})(i,j_{2})}| i \in I , \bar{Y}^{(d)}_{j_{1},j_{2}} \in q_{0} \right\} \cup \left\{Z^{(d)}_{(i,j_{1})(i,j_{2})}|i \in I , Y^{(d)}_{(i_{1},i_{2})} \in \tilde{q}_{*}\right\}     \nonumber\\
            \tilde{p} \times J &= \left\{\bar{Z}^{(c+n+1)}_{(i_{1},j)(i_{2},j)}|\bar{X}_{(i_{1}i_{2})}^{(c)} \in p, j \in J\right\}\nonumber \\
            p \times \tilde{q}_{0} &= \left\{\bar{Z}^{(c+d)}_{(i_{1},j_{1})(i_{2},j_{2})}|\bar{X}_{(i_{1}i_{2})}^{(c)} \in p, \bar{Y}_{j_{1}j_{2}}^{(d)} \in \tilde{q}_{0}\right\} \nonumber\\
            \cancel{p} \times \tilde{p}_{*} &= \left\{Z^{(c+d+1)}_{(i_{1},j_{1})(i_{2},j_{2})}|X^{(c)}_{i_{1}i_{2}} \in \cancel{p} , Y^{(d)}_{j_{1}j_{2}}\in \tilde{q}_{*}\right\}
        \end{align}  

         It is clear that with these definitions both $\tilde{p}\times q$ and $p \times \tilde{q}$ contain either the field or its conjugate for every field in $P_{p} \times Q_{q}$. We will now show that the fields in them also cover every term in the superpotential exactly once. 

We begin with $\tilde{p}\times q$ and consider $\mathcal{W}_P$, $\mathcal{W}_Q$ and $\mathcal{W}_C$ and $\mathcal{W}_{PQ}$ separately. Starting with $\mathcal{W}_P$ let us consider a term $T_{P}$ in the superpotential of $P$. This term gives rise to a number of terms in $\mathcal{W}_P$ as shown in \eref{potential_w_i_p_product} and \eref{potential_w_y_p_product}. Since $\tilde{p}$ is a perfect matching of $P$, then $T_{P}$ contains exactly one field from $\tilde{p}$. There are three possibilities for how such field appears in a term of $\mathcal{W}_P$ descending from $T_{P}$:
        \begin{itemize}
            \item 
                It gets replaced by its product with a node of $Q$. The resulting field is in $\tilde{p}\times J$ so this term is covered exactly once by $\tilde{p}\times q$.
            \item
                This field is common to $\tilde{p}$ and $p$ and gets replaced by its product with a field in $q$. The result is a field in $\tilde{p}_{0} \times q$.
            \item
                This field is in $\tilde{p}$ but not in $p$ and gets replaced by its product with a field not in $q$. The result is a field in $\tilde{p}_{*}\times \cancel{q}$. 
        \end{itemize}  
We conclude that in the three cases the field in $\tilde{p}$ that covers the term $T_{P}$ gives rise to exactly the field in a term descending from $T_{P}$ that is in $\tilde{p}\times q$. 

Similarly, for $\mathcal{W}_Q$ we consider the terms in it descending from $T_{Q}$. Such a term in $\mathcal{W}_Q$ always contains a field with one of its parents in $q$. There are three cases for what happens to this field in a term coming from $T_{Q}$:
        \begin{itemize}
            \item
                It gets replaced by its product with a node $i$ of $P$. The resulting field is in $I \times q$ so $\tilde{p} \times q$ covers this term exactly once.
            \item
                It gets replaced by its product with a field that is common to $p$ and $q$. In this case, this replacement is in $\tilde{p}_{0}\times q$ so $\tilde{p} \times q$ again covers this term once.
            \item
                It gets replaced by its product with $\bar{X}^{(m-c)}_{ii^{\prime}}$, a field in $p$ that is not in $\tilde{p}$. Unlike the previous case this replacement is not in $\tilde{p}_{0} \times q$. Since $\bar{X}^{(m-c)}_{ii^{\prime}}$ is not in $\tilde{p}$, its conjugate $X^{(c)}_{i^{\prime} i}$ is in $\tilde{p}$. As \eref{potential_w_x_q_product} shows, such a term also contains another field that comes from the product of $X^{(c)}_{i^{\prime} i}$ with a field not in $q$. This field is in $\tilde{p}_{*} \times \cancel{q}$ and hence $\tilde{p}\times q$ covers this term exactly once.
        \end{itemize}   
      
Let us now show that $\tilde{p}\times q$ covers every term in $\mathcal{W}_C$ exactly once. For this we inspect \eref{cubic_terms} and consider the following cases:
        \begin{itemize}
            \item
                If $X^{(c)}_{i^{\prime}i}$ is in $\tilde{p}$ then both $Z^{(c)}_{(i^{\prime},j)(i,j)}$ and $Z^{(c)}_{(i^{\prime},j^{\prime})(i,j^{\prime})}$ are in $\tilde{p}\times J$. Therefore, in this case $\tilde{p}\times q$ covers the two terms in \eref{cubic_terms} exactly once.
            \item
                If $X^{(c)}_{i^{\prime}i}$ is not $\tilde{p}$ then $\bar{X}^{(m-c)}_{ii^{\prime}}$ is in $\tilde{p}$ and hence in $\tilde{p}_{0}$. As a result $\bar{Z}^{(m+n-c-d)}_{(i_{1},i_{2})(j_{1},j_{2})}$ is in $\tilde{p}_{0}\times q$ and in this case $\tilde{p}\times q$ also covers the two terms in \eref{cubic_terms} exactly once. 
        \end{itemize}
        
Finally, let us focus on $\mathcal{W}_{PQ}$. A term in $\mathcal{W}_{PQ}$ has $T_{P}$ and $T_{Q}$ as parents. The field in $\tilde{p}$ that covers $T_{P}$ gives rise to exactly one field that is in $\tilde{p}\times q$ and covers this term. 

This completes our proof that $\tilde{p}\times q$ is a perfect matching. The same argument, exchanging the roles of $P$ and $Q$ along with $p$ and $q$, shows that $p \times \tilde{q}$ is also a perfect matching. It is important to note that we cannot use this process to construct $\tilde{p}\times \tilde{q}$ for arbitrary perfect matchings $\tilde{p}$ of $P$ and $\tilde{q}$ of $Q$. We must have either $\tilde{p} = p$ or $\tilde{q} = q$. This is consistent with the fact that $T_{\mathrm{CY}_{m+2}}$ is embedded in the plane spanned by the first $m+1$ coordinates with the last $n+1$ coordinates fixed to $v_{0}$. Similarly this also realizes the fact that $T_{\mathrm{CY}_{n+2}}$ is embedded in the plane spanned by the last $n+1$ coordinates with the first $m+1$ coordinates fixed to $u_{0}$. These positions for the perfect matchings give rise to the expected toric diagram.

Generically, the perfect matchings we have described are not all the perfect matchings of $P_{p} \times Q_{q}$. First, the final theory might have additional perfect matchings for the same points in $T_{\mathrm{CY}_{m+n+3}}$. Moreover, there might be new points in the toric diagram, which is the convex hull of the points corresponding to the perfect matchings we have constructed (see \fref{product_toric_diagram} for an example). Perfect matchings associated to these points are generated but do not descend from a pair of perfect matchings $\tilde{p}$ of $P$ and $\tilde{q}$ of $Q$.

\bibliographystyle{JHEP}
\bibliography{mybib}

\end{document}

%% file: pref.tex
\newcommand{\be}{\begin{equation}}
\newcommand{\ee}{\end{equation}}
\newcommand{\beq}{\begin{equation}}
\newcommand{\beql}[1]{\begin{equation}\label{#1}}
\newcommand{\eeq}{\end{equation}}
\newcommand{\ba}{\begin{array}}
\newcommand{\ea}{\end{array}}
\newcommand{\bea}{\begin{eqnarray}}
\newcommand{\beal}[1]{\begin{eqnarray}\label{#1}}
\newcommand{\eea}{\end{eqnarray}}
\newcommand{\ben}{\begin{enumerate}}
\newcommand{\een}{\end{enumerate}}
\newcommand{\bean}{\begin{eqnarray*}}
\newcommand{\eean}{\end{eqnarray*}}
\newcommand{\eref}[1]{(\ref{#1})}
\newcommand{\sref}[1]{\S\ref{#1}}

\newcommand{\fref}[1]{Figure \ref{#1}}
\newcommand{\btab}[1]{\begin{tabular}{#1}}
\newcommand{\etab}{\end{tabular}}

\def \Black {\pdfsetcolor{0 0 0 1}}

\newcommand{\comment}[1]{}

\newcommand{\CN}{{\cal N}}

\newcommand{\qed}{\nobreak \ifvmode \relax \else
      \ifdim\lastskip<1.5em \hskip-\lastskip
      \hskip1.5em plus0em minus0.5em \fi \nobreak
      \vrule height0.75em width0.5em depth0.25em\fi}

%% file: paper.bbl
\providecommand{\href}[2]{#2}\begingroup\raggedright\begin{thebibliography}{10}

\bibitem{Aldazabal:2000sa}
G.~Aldazabal, L.~E. Ibanez, F.~Quevedo and A.~Uranga, \emph{{D-branes at
  singularities: A Bottom up approach to the string embedding of the standard
  model}}, \href{http://dx.doi.org/10.1088/1126-6708/2000/08/002}{\emph{JHEP}
  {\bf 08} (2000) 002}, [\href{http://arxiv.org/abs/hep-th/0005067}{{\tt
  hep-th/0005067}}].

\bibitem{Berenstein:2001nk}
D.~Berenstein, V.~Jejjala and R.~G. Leigh, \emph{{The Standard model on a
  D-brane}},
  \href{http://dx.doi.org/10.1103/PhysRevLett.88.071602}{\emph{Phys.\ Rev.\
  Lett.} {\bf 88} (2002) 071602},
  [\href{http://arxiv.org/abs/hep-ph/0105042}{{\tt hep-ph/0105042}}].

\bibitem{Verlinde:2005jr}
H.~Verlinde and M.~Wijnholt, \emph{{Building the standard model on a
  D3-brane}},
  \href{http://dx.doi.org/10.1088/1126-6708/2007/01/106}{\emph{JHEP} {\bf 0701}
  (2007) 106}, [\href{http://arxiv.org/abs/hep-th/0508089}{{\tt
  hep-th/0508089}}].

\bibitem{Buican:2006sn}
M.~Buican, D.~Malyshev, D.~R. Morrison, H.~Verlinde and M.~Wijnholt,
  \emph{{D-branes at Singularities, Compactification, and Hypercharge}},
  \href{http://dx.doi.org/10.1088/1126-6708/2007/01/107}{\emph{JHEP} {\bf 01}
  (2007) 107}, [\href{http://arxiv.org/abs/hep-th/0610007}{{\tt
  hep-th/0610007}}].

\bibitem{Maldacena:1997re}
J.~M. Maldacena, \emph{{The Large N limit of superconformal field theories and
  supergravity}}, \href{http://dx.doi.org/10.1023/A:1026654312961}{\emph{Int.\
  J.\ Theor.\ Phys.} {\bf 38} (1999) 1113--1133},
  [\href{http://arxiv.org/abs/hep-th/9711200}{{\tt hep-th/9711200}}].

\bibitem{Gubser:1998bc}
S.~Gubser, I.~R. Klebanov and A.~M. Polyakov, \emph{{Gauge theory correlators
  from noncritical string theory}},
  \href{http://dx.doi.org/10.1016/S0370-2693(98)00377-3}{\emph{Phys.\ Lett.\ B}
  {\bf 428} (1998) 105--114}, [\href{http://arxiv.org/abs/hep-th/9802109}{{\tt
  hep-th/9802109}}].

\bibitem{Witten:1998qj}
E.~Witten, \emph{{Anti-de Sitter space and holography}},
  \href{http://dx.doi.org/10.4310/ATMP.1998.v2.n2.a2}{\emph{Adv.\ Theor.\
  Math.\ Phys.} {\bf 2} (1998) 253--291},
  [\href{http://arxiv.org/abs/hep-th/9802150}{{\tt hep-th/9802150}}].

\bibitem{Morrison:1998cs}
D.~R. Morrison and M.~R. Plesser, \emph{{Nonspherical horizons. 1.}},
  {\emph{Adv.Theor.Math.Phys.} {\bf 3} (1999) 1--81},
  [\href{http://arxiv.org/abs/hep-th/9810201}{{\tt hep-th/9810201}}].

\bibitem{Beasley:1999uz}
C.~Beasley, B.~R. Greene, C.~Lazaroiu and M.~Plesser, \emph{{D3-branes on
  partial resolutions of Abelian quotient singularities of Calabi-Yau
  threefolds}},
  \href{http://dx.doi.org/10.1016/S0550-3213(99)00646-X}{\emph{Nucl.Phys.} {\bf
  B566} (2000) 599--640}, [\href{http://arxiv.org/abs/hep-th/9907186}{{\tt
  hep-th/9907186}}].

\bibitem{Feng:2000mi}
B.~Feng, A.~Hanany and Y.-H. He, \emph{{D-brane gauge theories from toric
  singularities and toric duality}},
  \href{http://dx.doi.org/10.1016/S0550-3213(00)00699-4}{\emph{Nucl. Phys.}
  {\bf B595} (2001) 165--200}, [\href{http://arxiv.org/abs/hep-th/0003085}{{\tt
  hep-th/0003085}}].

\bibitem{Beasley:2001zp}
C.~E. Beasley and M.~R. Plesser, \emph{{Toric duality is Seiberg duality}},
  \href{http://dx.doi.org/10.1088/1126-6708/2001/12/001}{\emph{JHEP} {\bf 0112}
  (2001) 001}, [\href{http://arxiv.org/abs/hep-th/0109053}{{\tt
  hep-th/0109053}}].

\bibitem{Feng:2001xr}
B.~Feng, A.~Hanany and Y.-H. He, \emph{{Phase structure of D-brane gauge
  theories and toric duality}},
  \href{http://dx.doi.org/10.1088/1126-6708/2001/08/040}{\emph{JHEP} {\bf 08}
  (2001) 040}, [\href{http://arxiv.org/abs/hep-th/0104259}{{\tt
  hep-th/0104259}}].

\bibitem{Feng:2001bn}
B.~Feng, A.~Hanany, Y.-H. He and A.~M. Uranga, \emph{{Toric duality as Seiberg
  duality and brane diamonds}},
  \href{http://dx.doi.org/10.1088/1126-6708/2001/12/035}{\emph{JHEP} {\bf 12}
  (2001) 035}, [\href{http://arxiv.org/abs/hep-th/0109063}{{\tt
  hep-th/0109063}}].

\bibitem{Feng:2002zw}
B.~Feng, S.~Franco, A.~Hanany and Y.-H. He, \emph{{Symmetries of toric
  duality}}, \href{http://dx.doi.org/10.1088/1126-6708/2002/12/076}{\emph{JHEP}
  {\bf 12} (2002) 076}, [\href{http://arxiv.org/abs/hep-th/0205144}{{\tt
  hep-th/0205144}}].

\bibitem{Wijnholt:2002qz}
M.~Wijnholt, \emph{{Large volume perspective on branes at singularities}},
  \href{http://dx.doi.org/10.4310/ATMP.2003.v7.n6.a6}{\emph{Adv. Theor. Math.
  Phys.} {\bf 7} (2003) 1117--1153},
  [\href{http://arxiv.org/abs/hep-th/0212021}{{\tt hep-th/0212021}}].

\bibitem{Benvenuti:2004dy}
S.~Benvenuti, S.~Franco, A.~Hanany, D.~Martelli and J.~Sparks, \emph{{An
  infinite family of superconformal quiver gauge theories with Sasaki-Einstein
  duals}}, \href{http://dx.doi.org/10.1088/1126-6708/2005/06/064}{\emph{JHEP}
  {\bf 06} (2005) 064}, [\href{http://arxiv.org/abs/hep-th/0411264}{{\tt
  hep-th/0411264}}].

\bibitem{Franco:2005rj}
S.~Franco, A.~Hanany, K.~D. Kennaway, D.~Vegh and B.~Wecht, \emph{{Brane Dimers
  and Quiver Gauge Theories}},
  \href{http://dx.doi.org/10.1088/1126-6708/2006/01/096}{\emph{JHEP} {\bf 01}
  (2006) 096}, [\href{http://arxiv.org/abs/hep-th/0504110}{{\tt
  hep-th/0504110}}].

\bibitem{Benvenuti:2005ja}
S.~Benvenuti and M.~Kruczenski, \emph{{From Sasaki-Einstein spaces to quivers
  via BPS geodesics: L**p,q|r}},
  \href{http://dx.doi.org/10.1088/1126-6708/2006/04/033}{\emph{JHEP} {\bf 04}
  (2006) 033}, [\href{http://arxiv.org/abs/hep-th/0505206}{{\tt
  hep-th/0505206}}].

\bibitem{Franco:2005sm}
S.~Franco et~al., \emph{{Gauge theories from toric geometry and brane
  tilings}}, \href{http://dx.doi.org/10.1088/1126-6708/2006/01/128}{\emph{JHEP}
  {\bf 01} (2006) 128}, [\href{http://arxiv.org/abs/hep-th/0505211}{{\tt
  hep-th/0505211}}].

\bibitem{Butti:2005sw}
A.~Butti, D.~Forcella and A.~Zaffaroni, \emph{{The Dual superconformal theory
  for L**pqr manifolds}},
  \href{http://dx.doi.org/10.1088/1126-6708/2005/09/018}{\emph{JHEP} {\bf 09}
  (2005) 018}, [\href{http://arxiv.org/abs/hep-th/0505220}{{\tt
  hep-th/0505220}}].

\bibitem{Aspinwall:2008jk}
P.~S. Aspinwall, \emph{{D-Branes on Toric Calabi-Yau Varieties}},
  \href{http://arxiv.org/abs/0806.2612}{{\tt 0806.2612}}.

\bibitem{lam2014calabi}
Y.~T. Lam, \emph{Calabi-yau categories and quivers with superpotential},
  {\emph{PhD thesis, University of Oxford} (2014) }.

\bibitem{Franco:2017lpa}
S.~Franco and G.~Musiker, \emph{{Higher Cluster Categories and QFT Dualities}},
  \href{http://dx.doi.org/10.1103/PhysRevD.98.046021}{\emph{Phys. Rev.} {\bf
  D98} (2018) 046021}, [\href{http://arxiv.org/abs/1711.01270}{{\tt
  1711.01270}}].

\bibitem{Closset:2018axq}
C.~Closset, S.~Franco, J.~Guo and A.~Hasan, \emph{{Graded quivers and B-branes
  at Calabi-Yau singularities}},
  \href{http://dx.doi.org/10.1007/JHEP03(2019)053}{\emph{JHEP} {\bf 03} (2019)
  053}, [\href{http://arxiv.org/abs/1811.07016}{{\tt 1811.07016}}].

\bibitem{Hanany:2005ve}
A.~Hanany and K.~D. Kennaway, \emph{{Dimer models and toric diagrams}},
  \href{http://arxiv.org/abs/hep-th/0503149}{{\tt hep-th/0503149}}.

\bibitem{Franco:2015tna}
S.~Franco, D.~Ghim, S.~Lee, R.-K. Seong and D.~Yokoyama, \emph{{2d (0,2) Quiver
  Gauge Theories and D-Branes}},
  \href{http://dx.doi.org/10.1007/JHEP09(2015)072}{\emph{JHEP} {\bf 09} (2015)
  072}, [\href{http://arxiv.org/abs/1506.03818}{{\tt 1506.03818}}].

\bibitem{Franco:2015tya}
S.~Franco, S.~Lee and R.-K. Seong, \emph{{Brane Brick Models, Toric Calabi-Yau
  4-Folds and 2d (0,2) Quivers}},
  \href{http://dx.doi.org/10.1007/JHEP02(2016)047}{\emph{JHEP} {\bf 02} (2016)
  047}, [\href{http://arxiv.org/abs/1510.01744}{{\tt 1510.01744}}].

\bibitem{Franco:2016nwv}
S.~Franco, S.~Lee and R.-K. Seong, \emph{{Brane brick models and 2d (0, 2)
  triality}}, \href{http://dx.doi.org/10.1007/JHEP05(2016)020}{\emph{JHEP} {\bf
  05} (2016) 020}, [\href{http://arxiv.org/abs/1602.01834}{{\tt 1602.01834}}].

\bibitem{Franco:2016qxh}
S.~Franco, S.~Lee, R.-K. Seong and C.~Vafa, \emph{{Brane Brick Models in the
  Mirror}}, \href{http://dx.doi.org/10.1007/JHEP02(2017)106}{\emph{JHEP} {\bf
  02} (2017) 106}, [\href{http://arxiv.org/abs/1609.01723}{{\tt 1609.01723}}].

\bibitem{Franco:2016tcm}
S.~Franco, S.~Lee, R.-K. Seong and C.~Vafa, \emph{{Quadrality for
  Supersymmetric Matrix Models}},
  \href{http://dx.doi.org/10.1007/JHEP07(2017)053}{\emph{JHEP} {\bf 07} (2017)
  053}, [\href{http://arxiv.org/abs/1612.06859}{{\tt 1612.06859}}].

\bibitem{Franco:2017cjj}
S.~Franco, D.~Ghim, S.~Lee and R.-K. Seong, \emph{{Elliptic Genera of 2d (0,2)
  Gauge Theories from Brane Brick Models}},
  \href{http://dx.doi.org/10.1007/JHEP06(2017)068}{\emph{JHEP} {\bf 06} (2017)
  068}, [\href{http://arxiv.org/abs/1702.02948}{{\tt 1702.02948}}].

\bibitem{Franco:2019bmx}
S.~Franco and A.~Hasan, \emph{{Graded Quivers, Generalized Dimer Models and
  Toric Geometry}},
  \href{http://dx.doi.org/10.1007/JHEP11(2019)104}{\emph{JHEP} {\bf 11} (2019)
  104}, [\href{http://arxiv.org/abs/1904.07954}{{\tt 1904.07954}}].

\bibitem{Franco:2016fxm}
S.~Franco, S.~Lee and R.-K. Seong, \emph{{Orbifold Reduction and 2d (0,2) Gauge
  Theories}}, \href{http://dx.doi.org/10.1007/JHEP03(2017)016}{\emph{JHEP} {\bf
  03} (2017) 016}, [\href{http://arxiv.org/abs/1609.07144}{{\tt 1609.07144}}].

\bibitem{Franco:2018qsc}
S.~Franco and A.~Hasan, \emph{{$3d$ Printing of $2d$ $\mathcal{N}=(0,2)$ Gauge
  Theories}}, \href{http://dx.doi.org/10.1007/JHEP05(2018)082}{\emph{JHEP} {\bf
  05} (2018) 082}, [\href{http://arxiv.org/abs/1801.00799}{{\tt 1801.00799}}].

\bibitem{Seiberg:1994pq}
N.~Seiberg, \emph{{Electric - magnetic duality in supersymmetric nonAbelian
  gauge theories}},
  \href{http://dx.doi.org/10.1016/0550-3213(94)00023-8}{\emph{Nucl. Phys.} {\bf
  B435} (1995) 129--146}, [\href{http://arxiv.org/abs/hep-th/9411149}{{\tt
  hep-th/9411149}}].

\bibitem{Gadde:2013lxa}
A.~Gadde, S.~Gukov and P.~Putrov, \emph{{(0, 2) trialities}},
  \href{http://dx.doi.org/10.1007/JHEP03(2014)076}{\emph{JHEP} {\bf 03} (2014)
  076}, [\href{http://arxiv.org/abs/1310.0818}{{\tt 1310.0818}}].

\bibitem{Klebanov:1998hh}
I.~R. Klebanov and E.~Witten, \emph{{Superconformal field theory on
  three-branes at a Calabi-Yau singularity}},
  \href{http://dx.doi.org/10.1016/S0550-3213(98)00654-3}{\emph{Nucl.Phys.} {\bf
  B536} (1998) 199--218}, [\href{http://arxiv.org/abs/hep-th/9807080}{{\tt
  hep-th/9807080}}].

\bibitem{toappear0}
J.~Bao, S.~Franco, Y.-H.~H. He, E.~Hirst, G.~Musiker and Y.~Xiao, \emph{Machine
  Learning Quiver Gauge Theories, Work in progress}.

\end{thebibliography}\endgroup
